\documentclass[preprint,aps,amsmath,nofootinbib,tightenlines,showpacs]{revtex4}

\usepackage{graphicx, bm, epsfig, color, soul}
\usepackage[caption=false]{subfig}

\newcommand{\eps}{\epsilon}
\newcommand{\ra}{\rightarrow}
\newcommand{\diag}{\mathrm{diag}}
\newcommand{\V}{\Upsilon}
\newcommand{\amp}{\mathcal{A}}
\newcommand{\amps}{\mathcal{A}_{\rm{S}}}
\newcommand{\ampt}{\mathcal{A}_{\rm{T}}}
\newcommand{\logd}[2]{#2 \frac{d#1}{d#2}}
\newcommand{\la}{\Lambda}
\newcommand{\ve}{\varepsilon}
\newcommand{\bo}{\mathcal{O}}

\newcommand{\gb}{g_{\mathrm{B}}}
\newcommand{\sign}{\mathrm{sign\,}}

\begin{document}
\ifpdf{}
\DeclareGraphicsExtensions{.pdf, .jpg}
\else
\DeclareGraphicsExtensions{.eps, .jpg}
\fi

\preprint{CALT-TH-2018-015}

\title{Wilsonian effective field theory of 2D van Hove singularities}

\author{Anton Kapustin}
\email{kapustin@theory.caltech.edu}
\author{Tristan McKinney}
\email{tmckinney@caltech.edu}
\affiliation{Walter Burke Institute for Theoretical Physics, California Institute of Technology,
    Pasadena, CA 91125}
\author{Ira Z.~Rothstein}
\email{izr@andrew.cmu.edu}
\affiliation{Department of Physics, Carnegie Mellon University,
    Pittsburgh, PA 15213}
\date{\today}


\begin{abstract}
We study 2D fermions with a short-range interaction in the presence of
a van Hove singularity. It is shown that this system can be
consistently described by an effective field theory whose Fermi
surface is subdivided into regions as defined by a factorization
scale, and that the theory is renormalizable in the sense that all of
the counterterms are well defined in the IR limit. The theory has the
unusual feature that the renormalization group equation for the
coupling has an explicit dependence on the renormalization scale, much
as in theories of Wilson lines. In contrast to the case of a round
Fermi surface, there are multiple marginal interactions with
nontrivial RG flow. The Cooper instability remains strongest in the
BCS channel. We also show that the marginal Fermi liquid scenario for
the quasiparticle width is a robust consequence of the van Hove
singularity. Our results are universal in the sense that they do not
depend on the detailed properties of the Fermi surface away from the
singularity.
\end{abstract}

\pacs{71.10.Hf, 11.10.Gh, 71.18.+y, 74.20.-z}
\maketitle

\section{Introduction}

In the 1990s and early 2000s, extensive theoretical work was devoted
to the study of systems of fermions in 2D with the Fermi level close
to a van Hove singularity
\cite{GGA,Pattetal,Dzialosh,Gonzetal,ML,HSFR,ZanchiSchulz,HalbothMetzner,IKK}.
In such a system, the Fermi velocity vanishes at isolated points on
the Fermi surface which we will refer to as van Hove points. From a
theoretical standpoint, the van Hove singularity is one of the
simplest situations in which deviations from standard Fermi liquid
theory are expected. For example, the leading order computation of the
self-energy~\cite{GGA,Pattetal} shows that with a short-range
interaction, the width of the quasiparticles is linear in the energy,
a characteristic behavior of the Marginal Fermi Liquid
(MFL)~\cite{MFL}. Since the MFL paradigm has been proposed to explain
some peculiar properties of the normal state of high-$T_c$
superconductors, it was speculated that high-$T_c$ superconductors are
special due to their proximity to a van Hove
singularity~\cite{Pattetal,VHscenario}. While this scenario has fallen
out of favor, understanding the effect of van Hove singularities on
the Fermi liquid remains an important problem.

Most of the studies cited above focus on the 2D Hubbard model on a
square lattice at or near half-filling because of its relevance to
cuprate superconductors. In this model, the Fermi surface is
diamond-shaped and features two inequivalent van Hove points (i.e.
points where the Fermi velocity vanishes) as well as nesting. These
features complicate the analysis, and it is hard to disentangle the
effects of van Hove points and nesting. In this paper we study in
detail the case of a single van Hove point from the point of view of
Effective Field Theory (EFT). When applied to the case of a
nonsingular Fermi surface, the EFT approach explains the ubiquity of
both the Fermi liquid and BCS-type
superconductivity~\cite{BG,Polch,Weinberg,Shankar}.

As was previously noticed in~\cite{Gonzetal,HSFR}, the hyperbolic
dispersion law characteristic of electrons near a 2D van Hove point
leads to additional divergences not regulated by the Wilsonian cutoff
$\Lambda$, and necessitates the introduction of an additional
regulator which we take to be a Fermi velocity cutoff $\V$. $\V$ also
plays the role of a factorization scale which splits the Fermi surface
into two regions, $v_F>\V$ and $v_F<\V$, where two different power
counting schemes apply. In each region momenta are split into large
``label'' momenta and small ``residual'' components. Previous work on
the 2D van Hove singularity has been plagued by nonlocal divergences,
and a recent detailed study~\cite{GLK} concluded that the van Hove EFT
is not renormalizable when the Fermi level is exactly at the van Hove
singularity and has a very narrow range of applicability when the
Fermi level is close to it. However, as we show, when momenta are
properly power counted, all of the counterterms are independent of the
residual momenta in each respective region and therefore should be
considered local. Furthermore, the coupling in each region can only
depend upon the label momenta. The coupling can be assumed to be
independent of momenta only when all components of the momenta are
smaller than $\Lambda/\V $.

In the BCS channel, the RG equation for the coupling function
explicitly depends on the logarithm of the ratio of the Wilsonian
cutoff $\Lambda$ to the bandwidth $W$ and leads to the well known
double logarithmic running~\cite{Gonzetal,HSFR,HalbothMetzner}. This
dependence on the UV scale $W$ represents a form of UV/IR mixing and
has interesting consequences discussed below.

The situation is reminiscent of high energy scattering processes, such
as the Sudakov form factor, where the phase space of gauge bosons is
split into two regions which dominate the IR behavior. This splitting
leads to additional (rapidity) divergences which necessitate a new
regulator~\cite{jain} to distinguish between soft and collinear modes.
Summing contributions from these two sectors leads to a cancellation
of the regulator but, as in the present case, the cancellation leaves
behind a Cheshire log in the beta function. This in turn leads to
double logs in the renormalization group flow.

We utilize our results to study how a van Hove singularity modifies
the low energy behavior. In particular, we discuss the Cooper
instability and the range of applicability of the Marginal Fermi
Liquid scenario. We show that the Cooper instability is the strongest
in the BCS channel, as in the case of the circular Fermi surface, but
is also present for other kinematic configurations. This happens
because of additional marginal interactions which lead to a breakdown
of the Fermi liquid picture. We also show that a certain
generalization of the MFL scenario is a robust consequence of the van
Hove EFT.\@

\section{A toy model of a van Hove singularity}

In the 2D Hubbard model on a square lattice, there are two VH points
in the Brillouin zone: $p_{VH}=(0,\pi)$ and
$p_{VH}=(\pi,0)$. When the hopping parameters in the $x$ and
$y$ directions are not equal, the energies of these two points are
different. If the Fermi level is much closer to one than the other,
the effective field theory of a single VH singularity should apply. At
both of the VH points, $2p_{VH}\sim 0$. We assume there is a
unique VH point in the Brillouin zone and time-reversal ($T$)
symmetry, which takes $p\mapsto-p$, is present.
Therefore the singularity sits at the origin, a fixed point under $T$.

Such a scenario is realized by expanding the nearest-neighbor Hubbard
model Hamiltonian around the point $p=0$ to lowest order in
momentum components and assuming a zero-range interaction. The
resulting action is
\begin{equation} \label{action}
  S= \int dt\, d^2x
  \left[ \psi^\dagger(i\partial_t - \ve(-i\nabla)+\mu)\psi
    - \frac{g}{2}{(\psi^\dagger \psi)}^2\right],
\end{equation}
where the dispersion relation is
\begin{equation} \label{quadrdisp}
\ve(p)= p^2\equiv t_x p_x^2-t_y p_y^2
\end{equation}
and is unbounded from below. $p^2$ denotes the square of the 2D
vector $p$ with respect to the indefinite metric $\diag (t_x,-t_y)$.
It is convenient to set $t_x=t_y=1$ by rescaling $p_x$ and $p_y$, such
that metric becomes $\diag (1,-1)$, and absorbing a factor of
$1/{\sqrt {t_x t_y}}$ into $g$. If we regard $p_x,p_y$ as periodic
with period of order $k_B$, then $t_x,t_y$ are of order $W/k_B^2$,
where $W$ is the bandwidth.

As usual, all states with $\ve(p)<\mu$ are assumed to be occupied,
so in the free ($g=0$) limit the excitations of the system are
particles and holes, both with nonnegative energy. When the Fermi
level $\mu$ vanishes, the system has a discrete symmetry
$\psi\leftrightarrow\psi^\dagger,x\leftrightarrow y$ which exchanges
particles and holes. Furthermore, the quadratic dispersion relation
has $O(1,1)$ invariance, and the short-range interaction preserves
this symmetry. Also, for $\mu=0$, the action (\ref{action}) is
invariant under dilatations
\begin{equation}
\psi(t,x)\ra \lambda^{-1}\psi(\lambda^{2}t,\lambda x).
\end{equation}
Invariance with respect to Galilean boosts is spontaneously broken by
the Fermi sea for all values of $\mu$. As usual, the dilatation
symmetry is anomalous on the quantum level. Internal symmetries
include $U(1)$ particle-number symmetry and $SU(2)$ spin symmetry.

The interaction term in (\ref{action}) has zero range, and in momentum
space corresponds to a four-point vertex with no momentum dependence.
A naive justification for this simple ansatz is that any vertex with
more than four fermionic fields or polynomial momentum dependence is
irrelevant in the RG sense. Here we assume that under the RG
transformations the momenta scale as
\begin{equation}
p_x\ra \lambda p_x,\ p_y\ra \lambda p_y,
\end{equation}
so energy has scaling dimension $2$ and $\psi$ has scaling dimension
$1$. Then the chemical potential $\mu$ is relevant, the coupling $g$
is marginal, and more complicated interactions are irrelevant.

This naive argument is, as well known, incorrect, due to the fact that
momenta tangent to the Fermi surface should not scale under RG flow.
Moreover, the theory defined with a contact interaction,
eq.~\eqref{action}, is not consistent, as corrections to the
four-point function include nonlocal divergences which cannot be
absorbed into a renormalization of $g$~\cite{Gonzetal,HSFR}. These
divergences can be traced to the noncompactness of the Fermi surface.

\section{Setting up the Van Hove EFT}

To make the theory~\eqref{action} well defined, one must impose a
cutoff on momenta which will render the Fermi surface compact. This
cutoff is imposed in addition to the Wilsonian cutoff $|\ve({\bf
  p})|\leq\Lambda$. We assume $\Lambda$ is much smaller than the
bandwidth $W\sim{}k_B^2$. We also assume that $|\mu|\ll\Lambda$, so
the modes near the Fermi surface are not integrated out.

Let $\V$ denote this momentum cutoff. The largest possible value for
$\V$ is of order $k_B$, the size of the Brillouin zone, and thus it is
natural to assume that $\Lambda\ll\V^2$. The region
\begin{equation}
|p_\pm|\leq\V, \quad |p_+p_-|\leq\Lambda,
\end{equation}
where $p_+=p_x+p_y$ and $p_-=p_x-p_y$, will be called the VH
region.\footnote{Note that $\V$ breaks the $O(1,1)$ symmetry but
  preserves the particle-hole symmetry.} Within this region, the
dispersion law is
\begin{equation} \label{vh-dispersion}
  \ve = p_+p_-.
\end{equation}
We are using $\Lambda$ and $\Upsilon$ as both explicit regulators and
factorization scales. $\Upsilon$ has a natural value of order $V_F$,
the typical value of the Fermi velocity away from the VH points, but
it can also be chosen parametrically smaller. In any physical result
the dependence on $\V$ should cancel, since its role is merely to
separate the VH and NVH regions. On the other hand, in any physical
prediction $\Lambda$ is a placeholder for the RG scale.

The VH region is the part of the $\Lambda$-neighborhood of the Fermi
surface that is close to the saddle point. In this region, the
dispersion relation (\ref{vh-dispersion}) implies that if both
components of momentum are of the same order, then
$p_{\pm}\sim\sqrt{\Lambda}\ll\Upsilon$. In addition to these ``soft
modes,'' the VH region is populated by collinear and anticollinear
modes whose momenta scale as $(\Upsilon,\Lambda/\Upsilon)$ and
$(\Lambda/\Upsilon,\Upsilon)$ respectively.

\begin{figure}
  \epsfig{file=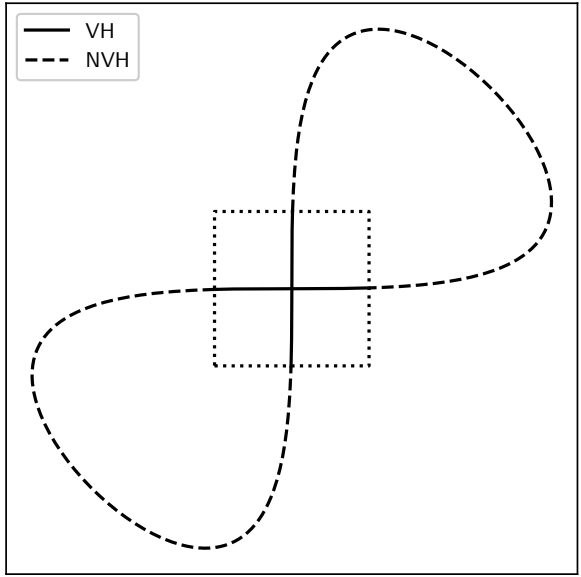, width=0.8\textwidth}
  \caption{An example of the division of the Fermi surface into van
    Hove and non-van Hove regions.\label{fig:vh-nvh-division}}
\end{figure}

The rest of the $\Lambda$-neighborhood of the Fermi surface will be
called the NVH region. Within this region, the dispersion law is the
standard
\begin{equation} \label{nvh-dispersion}
  \ve(p) = p_{\perp} v_{F}(p_{\parallel}),
\end{equation}
where $p_\perp/p_{\parallel}$ are normal/tangential to the Fermi
surface. We assume that the NVH region is ``featureless,'' in the
sense that the Fermi velocity does not change too much there, and that
it is free of nesting. The first assumption simply means that there
are no other van Hove singularities nearby, while the importance of
the second assumption will be discussed in
Section~\ref{sec:marginalFL}. Fig.~\eqref{fig:vh-nvh-division}
illustrates the division of a representative Fermi surface into the VH
and NVH regions.

In general, loop calculations involving modes from the VH region alone
will depend on $\V$ in such a way that the $\V\to{}\infty$ limit leads
to additional divergences. Thus, a sensible EFT must include both the
VH region and the NVH region. We use the term ``full theory'' for such
an EFT.\@ We make no assumptions about the shape of the Fermi surface
in the NVH region. As we will show below, our results are universal to
leading log accuracy in the sense that they only depend upon $V_F$,
the typical Fermi velocity in the NVH region, and not the detailed
shape of the Fermi surface. Therefore, our results apply to any system
with a VH singularity near the Fermi surface that is weakly coupled at
energies of order the bandwidth.

We will denote the fields annihilating electrons in the VH and NVH
regions $\psi_V$ and $\psi_N$ respectively. The interaction part of
the action is
\begin{equation} \label{vh-nvh-interaction}
S_{int}=\int dt \prod_{i=1}^4 d^2p_i \sum_{\alpha\beta\gamma\delta} g_{\alpha\beta\gamma\delta}\psi^\dagger_\alpha(p_1)\psi^\dagger_\beta(p_2)\psi_\gamma(p_3)\psi_\delta(p_4),
\end{equation}
where the indices $\alpha,\beta,\gamma,\delta$ take values $V$ and
$N$. In general, $g_{\alpha\beta\gamma\delta}$ can depend on the
momenta $p_{i}$ of the interacting modes. The couplings must match
onto each other as the VH modes approach the NVH region and vice
versa. For example, $g_{NNVV}$ must match onto $g_{VVVV}$ as $p_{1}$
and $p_{2}$ approach the VH region.

Naively, in light of the dispersion laws~\eqref{vh-dispersion}
and~\eqref{nvh-dispersion}, one might think that the coupling
functions in~\eqref{vh-nvh-interaction} should only depend on the
$p_{\parallel}$ of the NVH modes and that the only marginal
interaction between the VH modes should be a momentum-independent
constant. We will see in the next section that this is not
self-consistent: one-loop calculations imply that the couplings must
depend on momentum in a more generic manner. This is because when both
the rapidity cutoff $\V$ and the Wilsonian cutoff $\la$ are present, a
low momentum scale $\la/\V$ also plays a role. We will call $\la/\V$
the ultrasoft scale.

We can achieve some simplification by recalling that momentum and
energy conservation limits the interactions of the NVH modes to
special kinematic configurations~\cite{Shankar}. These configurations
correspond to forward scattering and back-to-back (BCS) scattering.
This implies that interactions between NVH modes and VH modes are of
two kinds: (1) forward scattering between a VH mode and an NVH mode
and (2) scattering of nearly back-to-back VH modes to nearly
back-to-back NVH modes and vice versa. As a result, the numbers of VH
and NVH particles are separately conserved.

\section{The one-loop beta function}\label{sec:one-loop}

\subsection{Generic kinematic configuration}

Consider the scattering of VH modes in a generic kinematic
configuration. Conservation of momentum implies the NVH modes will not
contribute. Thus tree-level interactions are described by a single
coupling function of three independent VH momenta. We would like to
determine how this function is renormalized.

It is enlightening to first assume that the coupling is a
momentum-independent constant, as naive power counting suggests. The
manner in which this assumption fails will show us how to
appropriately modify the theory.

\begin{figure}
  \epsfig{file=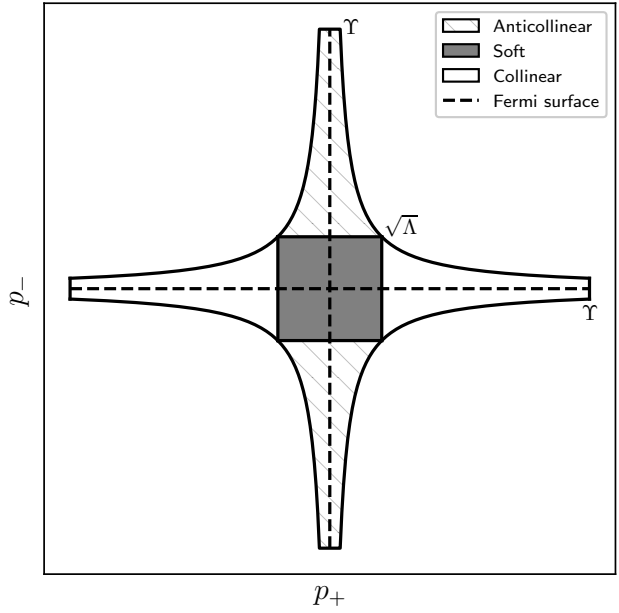, width=0.8\textwidth}
  \caption{Subdivision of the VH region.\label{fig:vh-regions}}
\end{figure}

We subdivide the VH region into three parts: the soft region, where
$p_{\pm}\sim \sqrt{\la}$; the collinear region, where $p_{+}\sim\V$
and $p_-\sim\frac{\Lambda}{\V}$; and the anticollinear region, where
$p_{-}\sim\V$, and $p_+\sim\frac{\Lambda}{\V}$.
Fig.~\ref{fig:vh-regions} illustrates the location of these
subregions. This separation is useful for categorizing the
contributions to the beta function. Since in this subsection the
kinematic configuration is assumed to be generic, the differences and
sums of external momenta are of the same order as the momenta
themselves.

\begin{figure}
  \subfloat[][]{\epsfig{file=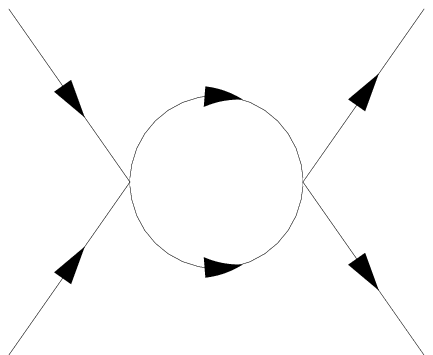, width=0.3\textwidth}}
  \hspace{0.2\textwidth}
  \subfloat[][]{\epsfig{file=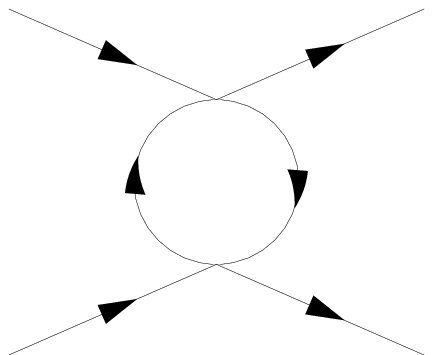, width=0.3\textwidth}}
  \caption{The diagram on the left/right is referred to as the
    $s$/$t$-channel diagram. Not shown is the $u$-channel diagram,
    which is given by interchanging the final state particles in the
    $t$-channel diagram.\label{fig:s-t-diagrams}}
\end{figure}

As usual, we have three diagrams at one-loop level, which we refer to
as $s$-channel ($\amps$), $t$-channel ($\ampt$), and $u$-channel
($\amp_{\rm{U}}$); see Fig.~\ref{fig:s-t-diagrams}. These three
diagrams depend on $K=p_{1}+p_{2}$, $Q=p_{1}-p_{3}$, and
$Q'=p_{1}-p_{4}$ respectively, and each contributes independently to
the beta function. The $u$-channel diagram is identical to the
$t$-channel diagram if we take $Q\leftrightarrow{}Q'$, so we focus on
the $t$- and $s$-channel diagrams.

We find (see Appendix) that any one-loop diagram where a collinear
external mode and an anticollinear external mode meet at a vertex
leads to a power-suppressed contribution to the beta function. This is
because the $K$ or $Q$ involved in the interaction always sets a large
energy scale which acts to suppress the associated diagram.

Generic $t$-channel diagrams that do not involve
collinear-anticollinear vertices make order-one contributions to the
beta function. For example, for a generic interaction between soft
modes,
\begin{equation} \label{ampt-generic}
  \logd{\ampt}{\la} = \frac{g^{2}}{4 \pi^{2}}
\end{equation}
plus power-suppressed terms. There are exceptions in certain special
kinematic configurations; see Section~\ref{sec:special-kinematics}.

The behavior of the $s$-channel diagrams is more complicated. Defining
\begin{equation} \label{define-ek}
  \ve_{K} = K_{+} K_{-},
\end{equation}
we find that generic the $s$-channel diagrams that do not involve
collinear-anticollinear vertices interpolate between being log
enhanced when $\ve_{K}\ll\la$ and order one when $\ve_{K}\sim\la$. As
an example, for generic interactions between soft modes,
\begin{equation} \label{amps-generic}
  \logd{\amps}{\la}
  = -\frac{g^{2}}{4 \pi^{2}}\log \left( \frac{ \la}{\ve_{K}} \right)
\end{equation}
plus suppressed terms. To avoid confusion, we note that $\ve_K$ is not
the net energy of the incoming particles.

\subsection{Special kinematic configurations}\label{sec:special-kinematics}

Eq.~\eqref{amps-generic} appears to imply that the beta function
diverges as $\ve_{K}$ approaches zero, thus necessitating the
existence of a nonlocal counterterm, which would mean the formalism
lacked a systematic power-counting scheme. However,
\eqref{amps-generic} does not apply in the $\ve_{K}\rightarrow 0$
limit. The divergent behavior is an unphysical artifact of taking the
van Hove region to be infinite in extent. If we take the rapidity
cutoff $\V$ into account, we find that when one component of $K$, say
$K_{-}$, satisfies
\begin{equation} \label{small-condition}
  |K_{-}| < \frac{\la}{\V},
\end{equation}
such as for an interaction between only collinear modes, then
\begin{equation} \label{amps-one-small}
  \logd{\amps}{\la}
  = -\frac{g^{2}}{4 \pi^{2}} \log \left( \frac{\V}{K_{+}} \right)
\end{equation}
plus order-one terms. If both components of $K$ are ultrasoft (i.e.
smaller in magnitude than $\la/\V$), we find to leading log order
\begin{equation} \label{amps-both-small}
  \logd{\amps}{\la}
   = -\frac{g^{2}}{4 \pi^{2}} \log{ \left( \frac{\V^{2}}{\la} \right) }.
\end{equation}
We can summarize the detailed behavior of the $s$-channel contribution
to the beta function in the following manner:\footnote{Note that for
  the s-channel diagram, taking $\ve_{K} \gtrsim \la$ is equivalent to
  injecting a large virtuality into the loop, which is formally
  outside the range of validity of the effective theory. The effects
  of such modes in intermediate states are properly accounted for in
  higher dimensional, power-suppressed, operators. This is consistent
  with the result in (\ref{amps-cases}).}
\begin{equation} \label{amps-cases}
  \logd{\amps}{\la} =
  \begin{cases}
    -\frac{g^{2}}{4 \pi^{2}} \log
    \left(
    \frac{\la}{\max{(K_{+}, \la/\V)}
      \max{(K_{-}, \la/\V)}} \right),
    & \ve_{K} \lesssim \la \\
    \bo(1) \times \frac{\la}{\ve_{K}} g^{2},
    & \ve_{K} \gtrsim \la.
  \end{cases}
\end{equation}

If $K_{+}\sim\V$, the log in~\eqref{amps-one-small} will not be large,
and hence the order-one ``corrections'' cannot be ignored. As a
result, the dependence on $K_{+}$ becomes complicated. Similarly, if
one component of $Q$ is large while the other is ultrasoft, the
$t$-channel diagram has a complicated dependence on the large
component (though unlike the $s$-channel diagram, it can never become
log enhanced). These cases are discussed in more detail in
Section~\ref{sec:marginalFL}. Finally, the $t$-channel contribution to
the beta function vanishes if both components of $Q$ are ultrasoft.

\subsection{Binning and leading-log behavior}

At first glance, the behavior of the beta function implied by the
above results is rather odd. The contribution from the $s$-channel
diagram in Eq.~\eqref{amps-cases} sometimes depends nonanalytically on
the momentum, and the functional form of the results change when the
components of $K$ or $Q$ pass a particular threshold (around the scale
$\la/\V$). Previous authors~\cite{GLK} have particularly regarded the
behavior of the $t$-channel diagram as a sign of unavoidable
nonlocality in the theory. However, as discussed in the next section,
similar behavior appears already for a circular Fermi surface, and is
dealt with using bins in momentum space of size $\Lambda/K_F$. This
notion of binning allows for a clear separation between large and
small momenta, and was previously used in the context of the theory of
non-relativistic heavy quarks~\cite{LMR}. Binning is also implicit in
the standard Fermi-surface RG~\cite{Shankar}. We apply the same method
here.

We divide momentum space into bins of size $\la/\V$, each with a label
momentum corresponding to the center of the bin and a residual
momentum, of order $\la/\V$, corresponding to the position within the
bin. The couplings are then indexed by the discrete label momenta, and
we can Taylor expand in the residual momenta. The beta function then
depends at leading order on the label momenta alone, and all results
are analytic in the residual momenta. The theory is therefore
renormalizable, although the couplings depends in an arbitrary way on
the label momenta. The same is true for a circular Fermi surface (see
the next section).

The non-analytic dependence on the net momentum implies that our
assumption of a momentum-independent coupling was inconsistent, and
the RG flow will generate dependence on the label momenta even for
modes within the soft region. While this complicated behavior
threatens the predictive power of the theory, we will see in
Section~\ref{sec:leading} that the enhancement of the beta function
for modes with small net momentum allows for several important
simplifications.

\section{Revisiting the round Fermi surface}\label{sec:round-section}

Let us revisit some old results involving a round Fermi surface. In
that context, previous
authors~\cite{BG}\cite{Shankar}\cite{Polch}\cite{Weinberg} found that
only certain coupling functions are present in the IR theory. In
particular, only forward scattering and interactions between
back-to-back particles (the BCS channel) are marginal, in the language
of effective field theory. Furthermore, these authors found that only
the BCS coupling is renormalized (that is, corrections to forward
scattering are power suppressed), and that the beta function for the
BCS interaction is in fact one-loop exact for generic round Fermi
surfaces~\cite{Shankar}\cite{Weinberg}.

These results hold in the limit where the Wilsonian cutoff $\Lambda$
on the energy of the modes included in the theory (or, in other words,
the ``width'' of modes around the Fermi surface) is taken to zero
while the size of Fermi surface itself is held fixed. For nonzero
$\Lambda$, near-forward and near-BCS scattering continue to be present
in the theory. To understand their role more precisely, let us
consider their contributions to the one-loop beta function.

We may parameterize a generic coupling function in terms $K$, $Q$, and
$Q'$, the same functions of the external momenta defined in
Section~\eqref{sec:one-loop}. As before, the $s$-, $t$-, and
$u$-channel diagrams (\cite{Shankar} calls these the BCS, ZS, and ZS'
diagrams), depend on $K$, $Q$, and $Q'$ respectively, and the $t$- and
$u$-channel diagrams are exchanged under $Q\leftrightarrow{}Q'$. BCS
scattering occurs for $K=0$ and forward scattering occurs when either
$Q$ or $Q'$ is zero.

It is straightforward to show that when any of these momenta are order
$K_{F}$ (the radius of the Fermi surface), the presence of the Fermi
surface forces the contribution from the corresponding one-loop
diagram to the beta function to be suppressed. For example, the log
derivative of the one-loop $s$-channel diagram is
\begin{equation} \label{amps forward}
  \logd{\amps}{\la}
  \sim \frac{\la}{v_{F} |K|} g^{2}
\end{equation}
when $|K|\approx K_{F}$. A similar statement holds for the $t$-channel
and $u$-channel diagrams.

From this point of view, the one-loop contributions are generically
power suppressed. The exceptional behavior occurs when $K$ (or $Q$ or
$Q'$) is of order $\la/v_{F}$. Unlike the case for large $K$ or $Q$,
the behavior qualitatively differs between the $s$ and $t$ channels.

\begin{figure}
  \epsfxsize=8cm\epsfbox{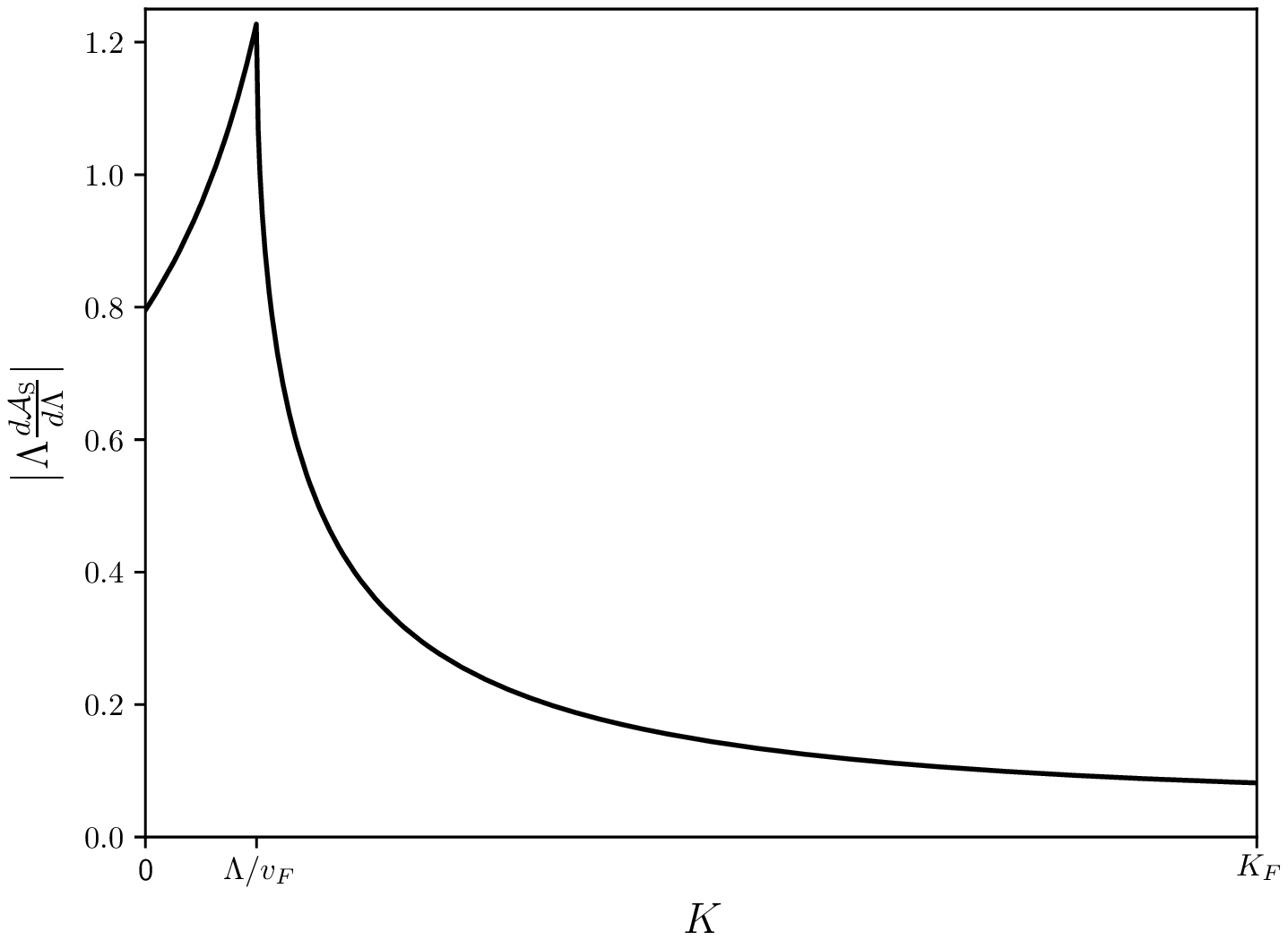}
  \epsfxsize=8cm\epsfbox{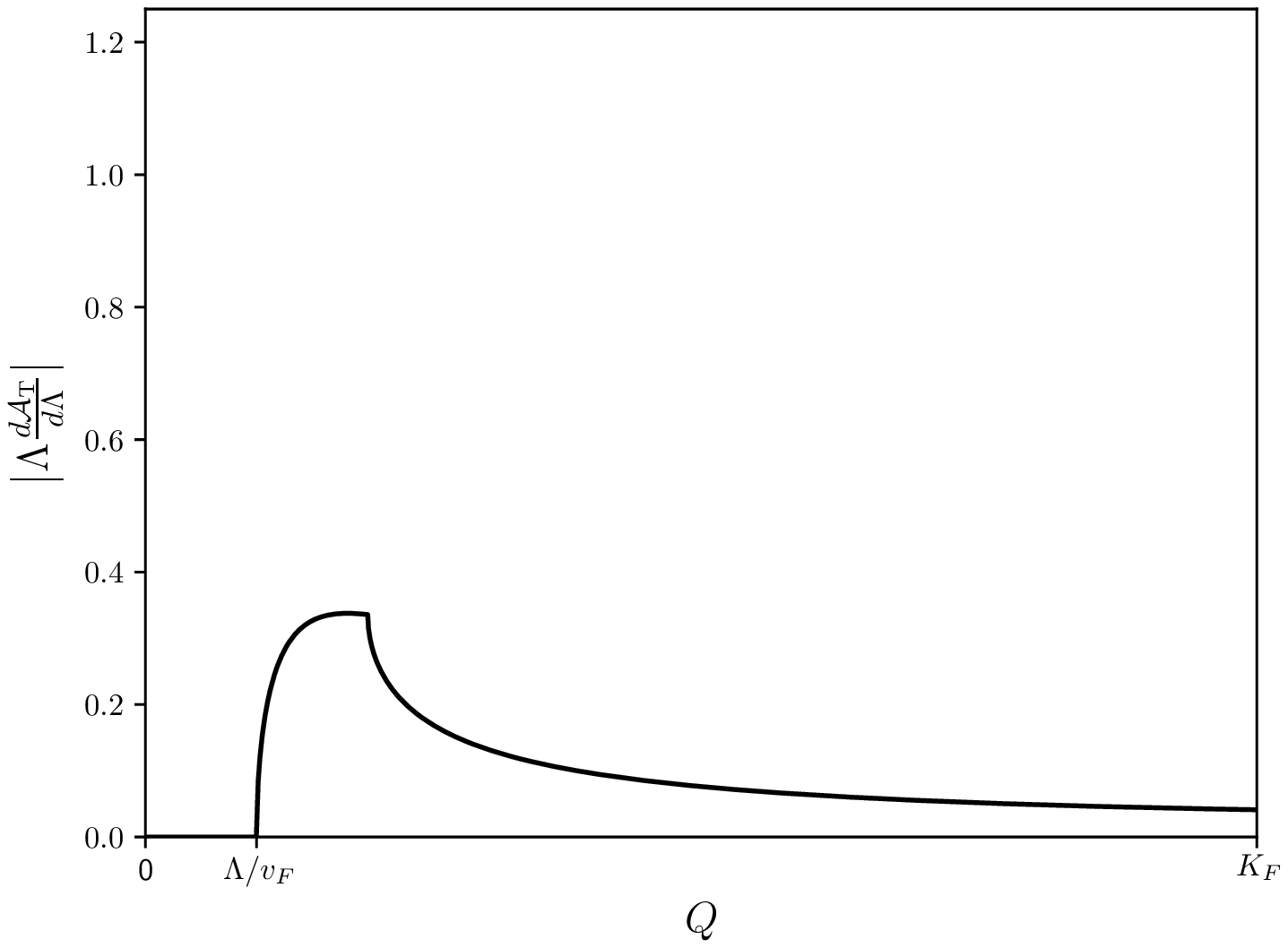}
  \caption{The log derivatives of the $s$- and $t$-channel diagrams
    measured in units of $g^{2}$ for a circular Fermi surface. We
    assume a constant coupling.\label{fig:round}}
\end{figure}

For the $t$-channel diagram to make a nonsuppressed contribution to
the beta function, the following must hold:
\begin{equation} \label{t condition}
  \frac{\la}{v_{F}} < |Q| < \bo(1) \times \frac{\la}{v_{F}}.
\end{equation}
Thus, there is a window of values where the contribution is nonzero,
and the position of the edges of this widow depend on $\la$. On the
other hand, for the $s$-channel diagram to make an unsuppressed
contribution to the beta function, $K$ must satisfy
\begin{equation} \label{s condition}
  |K| < \bo(1) \times \frac{\la}{v_{F}}.
\end{equation}
In particular, $K=0$ gives an order-one contribution while $Q=0$ does
not. Fig.~\ref{fig:round} demonstrates the behavior of the log
derivatives assuming a constant coupling.

This difference has a profound effect. In the course of the RG flow,
the condition that $K$ or $Q$ is order $\la/K_{F}$ changes, since we
take $\la$ to scale down. If $K$ is actually zero from the beginning,
there will always be an order-one contribution to the beta function,
and this condition is stable throughout the RG flow. This allows
attractive couplings with $K=0$ (the BCS channel) to become strong at
small $\Lambda$. On the other hand, the condition for the $t$-channel
diagram to give a unsuppressed contribution to the beta function is
not stable under the RG flow. Hence for {\it any\/} fixed $Q$, the $t$
channel only contributes to the beta function for a small period of RG
time.

In summary, the contribution to the beta function is power-suppressed
throughout the RG flow for generic (large) $K$ and $Q$. If $K$ or $Q$
is small enough, there are order-one contributions to the beta
function, but only for a short RG time. The only exception is ``true''
BCS scattering, where $|K|<\la/v_{F}$ throughout the flow. If we
assume the UV coupling is weak, this means the only coupling that can
be relevant to the ground state instability involves the BCS
configuration.

With this context, the startling results for the one-loop VH beta
function [Eqs.~\eqref{ampt-generic}-\eqref{amps-cases}] are less
surprising. Even with a round Fermi surface, the beta function, and
therefore the coupling, depends on $K$ and $Q$. This is even true for
the BCS coupling, which is generically a function of two angular
coordinates~\cite{Shankar,Weinberg} (playing the role of label
momenta) for noncircular Fermi surfaces.

Finally, the transition from zero contribution to the beta function
from the $t$-channel diagram to a finite contribution as we increase
$Q$ from zero is also present for the circular Fermi surface. The
major difference in the VH case is the long, flat section of the Fermi
surface, which guarantees that the window in $Q$ for which the $t$
channel is not power suppressed is larger than for a circular Fermi
surface. Fortunately, we will see that at least for certain
observables, we may once again neglect the contribution from the $t$
channel relative to the $s$ channel.

\section{The leading contribution at one loop}\label{sec:leading}

Section~\ref{sec:one-loop} demonstrates that at one loop, only the
$s$-channel diagram contains a logarithmic enhancement. Furthermore,
the largest possible contribution to the beta function occurs when
$K\simeq0$. This indicates that the kinematic configuration of
near-zero net momentum, the BCS channel,\footnote{In this context, the
  term ``BCS'' means back-to-back up to an ultrasoft momentum. Generic
  configurations of only ultrasoft modes therefore qualify as BCS.}
dominates the low-energy behavior of the theory.

With this in mind, assume the UV the dependence on the external
momenta is analytic. This condition will not be preserved under the
RG, because the $s$-channel introduces a nonanalytic dependence on the
net momentum $K$ in the four-point coupling. However, if we focus on
the BCS configuration we may ignore any nonanalytic dependence on the
other momenta to leading-log order.

In the following calculations, we sum the leading VH and NVH
contributions. While the precise form of the full results generically
depends on the detailed shape of the NVH portion of the Fermi surface,
the leading contribution is independent of these details. Instead,
this summing procedure turns out to be identical to taking the VH
results and replacing the cutoff $\V$ with $V_{F}$, its natural value.

Parameterize the BCS coupling as $\gb(p_{1}, p_{3})$, where $p_{1}$ is
the label momentum of one of the incoming pair of particles (the other
has label momentum $-p_{1}$) and $p_{3}$ is the label momentum of one
of the pair of outgoing particles. We find with logarithmic accuracy
(see Appendix)
\begin{equation} \label{bcs-beta}
  \la \frac{d \gb(p_{1}, p_{3})}{d \la}
  = \frac{1}{4 \pi^2} \gb(p_{1}, 0) \gb(0, p_{3})
  \log{\frac{V_{F}^{2}}{\la}}.
\end{equation}

An unusual feature of this equation is that the beta function has an
explicit dependence on $\Lambda$, as well as $V_F^2$. The latter can
be regarded as an energy scale of order of the bandwidth,
$V_F^2\sim{}W$. Thus the IR physics retains some information about the
UV scale $W$. 

The solution to~\eqref{bcs-beta} is
\begin{equation} \label{bcs-solved}
  \gb(p_{1}, p_{3}; \la)
    = \gb(p_{1}, p_{3}; \la_{0}) - 
    \left( \frac{1}{8 \pi^2} \right)
    \frac{\gb(p_{1}, 0; \la_{0}) \gb(0, p_{3}; \la_{0})
      \left(\log^2 \frac{V_F^2}{\la}
      - \log^2\frac{V_F^2}{\la_0}\right)}{1 +
      \frac{\gb(0, 0; \la_0)}{8\pi^2}
      \left(\log^2 \frac{V_F^2}{\la}
      - \log^2\frac{V_F^2}{\la_0}\right)}.
\end{equation}
The coupling in the vicinity of the van Hove singularity, $\gb(0,0)$,
plays a special role: it ``drives'' the RG for the other couplings,
and when it is attractive at the scale $\la_{0}$, it sets the one-loop
estimate of the strong-coupling scale,
\begin{equation} \label{Lambdas}
  \Lambda^{*}
  =V_F^2 \exp\left(-\sqrt {\log^2\frac{{V_F}^{2}}{\Lambda_0}
    +\frac{8\pi^2}{|\gb(0, 0; \la_{0})|}}\right).
\end{equation}

As in the ordinary BCS theory~\cite{BCStheory} the strong-coupling
scale is non-perturbative in $g(\la_{0})$. However, the usual
dependence of this scale on the microscopic parameters differs
from~\eqref{Lambdas}. While~\eqref{Lambdas} simplifies considerably if
we set $\Lambda_0=V_F^2\sim W$, this choice may be unphysical if the
van Hove EFT is obtained by integrating out some other degrees of
freedom at a scale below $W$. For example, if the short-range
interaction arises both from the screened Coulomb repulsion and the
phonon-mediated attraction, the van Hove EFT applies only up to energy
scales of order the Debye frequency $\omega_D$, which is usually much
smaller than the bandwidth $W$. Then the natural choice for
$\Lambda_0$ is $\omega_D$, and we have a hierarchy of scales
$V_F^2\simeq W\gg \omega_D$.

To understand some of the limitations of this formalism, consider the
amplitude (as opposed to the beta function) in the BCS configuration.
If we assume a momentum-independent BCS coupling, it is
straightforward to evaluate the one-loop amplitude with logarithmic
accuracy:
\begin{equation} \label{AVH}
  \amp_{\rm{BCS}}(E) =
  \frac{\gb^2}{8\pi^2}
  \left(\log^2 \frac{V_{F}^{2}}{\la}
  -\log^2 \frac{V_{F}^{2}}{E}
  - i\pi\log\frac{V_F^2}{E} \right),
\end{equation}
where we have kept only the leading terms in the real and imaginary
parts. Taking the log derivative of equation~\eqref{AVH} with respect
to $\la$ reproduces the beta function~\eqref{bcs-beta} for $\gb(0,0)$.
However, the imaginary part of the amplitude depends on
$\log\frac{V_F^2}{E}$. This large log is not resummed by the standard
beta function and indicates that something akin to the rapidity
renormalization group introduced in~\cite{jain} would be necessary to
resum subleading logs.

In the special case $\V^2=\Lambda$ our scheme in the VH region
resembles that of Ref.~\cite{ML}. In that work it is implicitly
assumed that $g$ is repulsive, and that $\Lambda$ can be taken as high
as the bandwidth, so that the NVH region is effectively absorbed into
the VH region. However, lowering $\Lambda$ then results in integrating
some low-energy modes and requires nonlocal counterterms.

\section{Higher-order renormalization}\label{sec:higherorder}

Let us discuss how higher-order corrections modify Eq.~\eqref{AVH}.
This is particularly important for the kinematic configuration with
zero net momentum which controls the Cooper instability. Since the
beta function at zero net momentum contains a logarithm of a large
ratio, $\log(V_F^2/\Lambda)$, one may wonder if the one-loop
computation is reliable in this kinematic configuration, or if one
needs to resum the logs in the beta function itself. We will call logs
containing $V_F^2$, such as $\log(V_F^2/E)$ or $\log(V_F^2/\Lambda)$,
rapidity logs. We want to estimate the contribution of higher rapidity
logs to the beta function at zero net momentum. 

\begin{figure}
  \epsfig{file=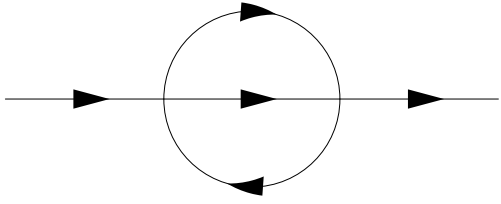, width=0.4\textwidth}
  \caption{The two-loop self-energy with finite imaginary
    part.\label{fig:sunrise}}
\end{figure}

We will limit ourselves to the analysis of 2-loop diagrams. We take $\V\sim V_F$, in which case
there are no large rapidity logs in non-VH loops. The renormalized
coupling $g$ is related to the bare coupling $g_b$ by
\begin{equation}
g_b=g Z_4 Z_2^{-2},
\end{equation}
where $Z_4$ is the renormalization factor for the particle-particle
four-point amplitude, and $Z_2$ is the wave function renormalization.
$Z_2$ is finite at one loop, and at two-loop order is determined from
the on-shell behavior of the self-energy diagram,
Fig.~\ref{fig:sunrise}, whose imaginary part is finite even without
the rapidity cut-off~\cite{GGA,Pattetal}, and therefore does not
contain rapidity logs.

\begin{figure}
  \subfloat[][]{\epsfig{file=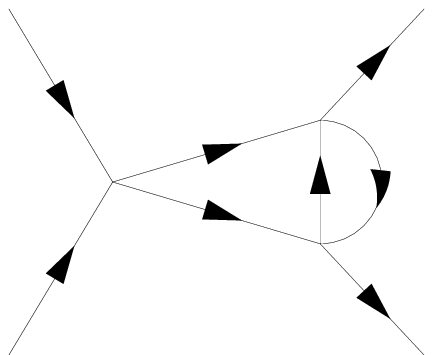,
      width=0.3\textwidth}\label{fig:two-loop-a}}
  \hspace{0.2\textwidth}
  \subfloat[][]{\epsfig{file=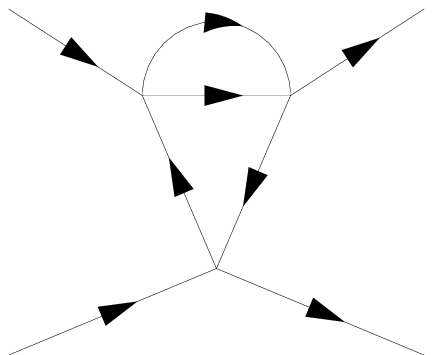,
      width=0.3\textwidth}\label{fig:two-loop-b}}
  \caption{Examples of two-loop contributions to the beta function.
    The diagrams with iterated loops are not shown. The diagram on the
    right can contribute a double rapidity log to the beta
    function.\label{fig:two-loop}}
\end{figure}

Two-loop contributions to $Z_4$ arise from diagrams such as in
Fig.~\ref{fig:two-loop}. (Iterations of one-loop diagrams do not
contribute since their infinities are removed by one-loop
counter-terms). Their contributions to the beta function can be
estimated using what we already know about the one-loop diagrams. For
example, the diagram Fig.~\ref{fig:two-loop-a} is the obtained from
the one-loop $s$-channel diagram by replacing one of the vertices with
the one-loop $t$-channel diagram. The latter does not contain rapidity
logs, so the contribution of the whole diagram to the beta function
should behave in the same way as that of the one-loop $s$-channel
diagram. In particular, it contains at most a single
$\log(V_F^2/\Lambda)$ at zero net momentum. The diagram
Fig.~\ref{fig:two-loop-b} can be regarded as a one-loop $t$-channel
diagram with one vertex replaced with a one-loop $s$-channel diagram.
The latter amplitude contains at most two rapidity logs, so the
contribution of Fig.~\ref{fig:two-loop-b} to the beta function
contains at most $\log^2(V_F^2/\Lambda)$. We conclude that with
logarithmic accuracy the two-loop beta function at zero net momentum
has the form
\begin{equation}
  \beta(g)=\frac{1}{4\pi^2} (g^2+C g^3)\log\frac{V_F^2}{\Lambda}
  + C' g^3 \log^{2}\frac{V_F^2}{\Lambda},
\end{equation}
where $C$ and $C'$ are constants.

Now we can see if the resummation of rapidity logs in the beta
function is necessary. Eq.~\eqref{bcs-solved} indicates that the
one-loop RG equations resum logs of the form $g
\log^2(V_F^2/\Lambda)$. Thus we are assuming that
$g\log^2(V_F^2/\Lambda)\lesssim1$, while $g\log(V_F^2/\Lambda)\ll1$.
This implies $g^3\log^{2}(V_F^2/\Lambda)$ is parametrically suppressed
relative to $g^2\log(V_F^2/\Lambda)$. We conjecture that this behavior persists at higher loops, in the sense that every extra power of $g$ is accompanied by at most a single rapidity log. If this is true, then resumming the
rapidity logs in the beta function will not will not change qualitative conclusions regarding the RG flow and the Cooper instability.

\section{The collinear region as a marginal Fermi liquid}\label{sec:marginalFL}

By definition, the collinear region is the part of the VH region where
$|p_+|$ is of order of the rapidity cutoff $\Upsilon$, while $|p_-|$
is less or equal than $\Lambda/\Upsilon$. The anticollinear region is
defined similarly, but with $p_+$ and $p_-$ exchanged. Each of the
following statements regarding the collinear region also applies to
the anticollinear region.

Everywhere in the collinear region, the Fermi velocity is nonzero.
Naively, one might conclude that this region is no different from the
NVH region. In particular, one might think that the usual Fermi
surface EFT~\cite{Shankar} applies both in the NVH and the collinear
region, but this is incorrect. To see why, recall that canonical Fermi
surface EFT predicts that all interactions (apart from forward and BCS
scattering) are irrelevant, and thus the quasiparticle width scales
like $E^2/v_F k_B$ for small $E$. In the NVH region, $v_F$ is of order
$W/k_B$, thus the Fermi liquid theory applies for $E$ much smaller
than $W$. But it is well known~\cite{Shankar} that additional marginal
interactions arise when a portion of the Fermi surface is related to
another portion of the Fermi surface by a translation in momentum
space (nesting). The translation vector $Q$ is called the nesting
vector. The collinear region is an extreme example of this, since the
Fermi surface is approximately invariant with respect to arbitrary
shifts with $Q=(Q_+,0)$. Following Wilczek and
Nayak~\cite{WilczekNayak}, we will refer to such a Fermi surface as
flat.

Wilczek and Nayak emphasized the failure of the Fermi liquid theory
for flat Fermi surfaces and proposed that the correct EFT for flat
Fermi surfaces is quasi-1D, with the component of momentum parallel to
the Fermi surface playing the role of a continuous label. In
particular, the four-fermion interaction is marginal for generic
combinations of momenta rather than irrelevant.

But there is also an important difference between the collinear region
and the model of interacting 1D fermions (the Luttinger model). In the
Luttinger model, the coupling is exactly marginal (has vanishing beta
function). This is most easily seen using bozonization, which turns
the Luttinger model into a free boson with a linear dispersion law.
The vanishing of the beta function does not apply to the EFT
describing the collinear region. The reason is that, unlike in the 1D
case, the Fermi velocity varies along the Fermi surface. For
definiteness, let us consider the collinear region and set $\mu=0$.
Then the ``small'' component of momentum is $p_-$, while the ``large''
one is $p_+$. If we treat $p_+$ as a continuous label, the Fermi
velocity is
\begin{equation}
v_F(p_+)=p_+.
\end{equation}
As long as we consider generic scattering events between particles for
which $p_{+}$ is $\bo(\V)$, the four-fermion coupling can be Taylor
expanded in $p_-$, but not in $p_+$. Thus the leading interaction term
\begin{equation}\label{collinearint}
  S_{int}
  = \int dt\int d^2 p_1 d^2 p_2 d^2 p_3
  \frac{1}{4} g(p_{1+},p_{2+},p_{3+})
  \eps^{\alpha_1 \alpha_2} \eps^{\alpha_3 \alpha_4}
  \psi_{\alpha_1}\psi_{\alpha_2}
  \psi^\dagger_{\alpha_3}\psi^\dagger_{\alpha_4}
\end{equation}
depends on a function of three real variables
$g(p_{1+},p_{2+},p_{3+})$ which we take to be spin independent. This
choice of spin structure for the interaction corresponds to the
spin-singlet coupling, which we will focus on here. Furthermore, we
take $g$ to be symmetric under $p_{1}\leftrightarrow{}p_{2}$ and
$p_{3}\leftrightarrow{}p_{1}+p_{2}-p_{3}$ independently, so the vertex
factor is
\begin{equation} \label{vertex-factor}
  i (\delta_{\alpha_1 \alpha_3} \delta_{\alpha_2 \alpha_4}
  -\delta_{\alpha_1 \alpha_4} \delta_{\alpha_2 \alpha_3})
  g(p_{1}, p_{2}, p_{3}).
\end{equation}

It is straightforward to compute the beta function for $g$. We find:
\begin{multline}\label{collinear-beta-final}
 \frac{dg(p_{1+}, p_{2+}, p_{3+})}{d\log\mu}
  = \frac{1}{2\pi^2}\int_{K}^{\V} d q\,
  \frac{g(p_{1+}, p_{2+}, q) g(q, K - q, p_{3+})}
       {2 q - K} \\
  + \frac{1}{8\pi^2 Q}\int_{Q-\min{(Q,\Upsilon)}}^{\min{(Q,\Upsilon)}} dq\,
  g(p_{1+}, q, p_{3+}) g(p_{2_{+}}, q+Q, p_{4+}) \\
  + \frac{1}{8\pi^2 Q'}\int_{Q'-\min{(Q',\Upsilon)}}^{\min{(Q',\Upsilon)}} dq\,
  g(p_{1+}, q, p_{4+}) g(p_{2_{+}}, q+Q', p_{3+})
\end{multline}
plus terms suppressed by $\ve(p)/\la$, where $p$ is one of the
external momenta. Here $K=p_{1+}+p_{2+}$ and $Q=p_{1+}-p_{3+}$ are
assumed to be positive, for definiteness. Even if we take $g$ to be
independent of the ``large'' components of momenta at some scale, the
RG evolution is nontrivial and introduces momentum dependence. At
higher orders we will also have to take into account the
renormalization of the Fermi velocity function $v_F(p_+)$. Finally, we
neglected the spin-triplet coupling. Even if it is set to zero in the
UV, it will be generated by radiative corrections, and thus a
renormalizable theory should have both couplings. The above
computation which takes into account only the spin-singlet coupling
merely illustrates our point that the beta functions are nonzero in
the collinear region.

\begin{figure}
  \epsfxsize=8cm\epsfbox{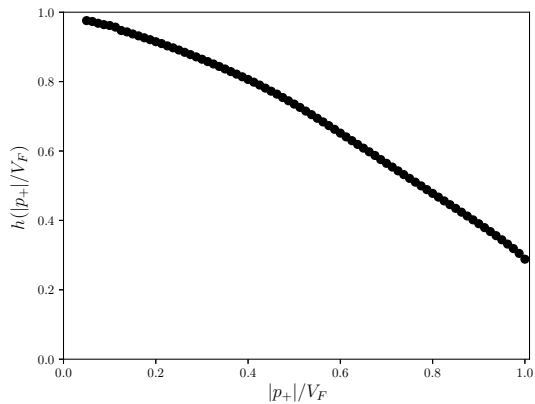}
  \caption{Numerical results for the dependence of $h$ on
    $|p_{+}|/V_{F}$ in units of $g^{2}$ assuming a constant coupling.
    $h$ is normalized to $g^{2}$ for
    $V_{F}\to\infty$.\label{fig:hfunction}}
\end{figure}

The EFT which includes only the collinear region is sufficient to
compute the width of the quasiparticle whose momentum is in the
collinear region, where $|p_{+}|$ is of order $\V$. If one formally
takes the limit $\V\ra\infty$ and assumes that the coupling $g$ is
independent of momenta, the leading-order computation can be performed
in the toy model and gives~\cite{Pattetal,GGA}:
\begin{equation}
\Gamma(E)\sim g^2 E.
\end{equation}
The linear dependence on $E$ follows from dimensional analysis and is
a hallmark of the marginal Fermi liquid~\cite{MFL}. The computation in
the toy model cannot be extended to higher orders, since it is not a
renormalizable theory. However, if we include the NVH region by
introducing the rapidity cutoff $\V=V_F$, dimensional analysis gives a
similar result:
\begin{equation} \label{width}
\Gamma(E)\sim h(|p_+|/V_F) E,
\end{equation}
where $V_F$ is the typical Fermi velocity in the NVH region. At
leading (two-loop) order the function $h(x)$ is of order $g^{2}$, but
is not a constant even if one assumes, for simplicity, that $g$ is a
constant. Evaluating the imaginary part of the self-energy diagram
(Fig.~\ref{fig:sunrise}) numerically, we find the result in
Fig.~\ref{fig:hfunction}.

The expression (\ref{width}) is valid provided we can neglect the
chemical potential $\mu$ which is a relevant coupling. Thus it holds
in the range $|\mu|\ll E\ll W$. The corrections are of several sorts.
The NVH region contribution is of order $E^2/W$, as usual. The
corrections from a nonzero $\mu$ are of order $\mu^2/E$. Finally,
higher orders in perturbation theory will give the function $h$ a weak
(logarithmic) dependence on $E$.

One of the defining properties of the MFL is that the quasiparticle
width, defined via the imaginary part of the on-shell self-energy, is
proportional to energy. The above arguments show that the Marginal
Fermi Liquid behavior~\cite{MFL} is a robust consequence of the
proximity to a van Hove singularity. On the other hand, the dependence
of the width on the ``large'' component of momentum can be nontrivial,
unlike in the simplest models of Marginal Fermi Liquids.

\section{Conclusions}

We have presented a systematic effective field theory description of
systems with a van Hove singularity. The formalism is valid to leading
power in an expansion in $E/W$ and generalizes the classic results in
\cite{Shankar,Polch}. We have shown that the theory is renormalizable
with all counterterms being local in the sense that they are finite in
the zero energy limit. That such a formalism exists had to be the case
given that any well-defined microscopic local theory must yield a
renormalizable description, if it is properly formulated. A crucial
ingredient in generating such a theory is the inclusion of all the
relevant modes on the Fermi surface. Given that the entire surface is
necessarily part of the IR description of the theory, it is not
surprising that focusing solely on one region leads to nonlocalities.

The EFT that we constructed depends on a coupling function
$g(k_1,k_2,k_3,-k_1-k_2,-k_3)$ that cannot be expanded in powers of
momenta (except when all momenta are ultrasoft). The appearance of an
arbitrary function of six variables makes the theory much less
predictive than the usual Fermi surface RG which has two marginal
couplings which depend on two variables each (for a 2D Fermi liquid).
Nevertheless, we showed that in the BCS channel the EFT can be greatly
simplified, provided we keep only logarithmically-enhanced terms. In
this channel, one is left with a single function of two variables
which satisfies a simple RG equation.

We have utilized our formalism to show that generic theories with van
Hove singularities will lead to Marginal Fermi Liquid behavior as
previously anticipated using toy models~\cite{GGA,Pattetal}. This
behavior arises in both the soft and collinear subsectors of the VH
region, the latter of which can constitute a considerable fraction of
the Fermi surface. Thus our conclusions disagree with~\cite{ML}, where
it was argued that for $E\gg\mu$ the Fermi liquid picture is valid.
Our treatment of the collinear region clarifies the physics of Fermi
surfaces with flat regions as discussed in~\cite{WilczekNayak}. We
also show that the running of the coupling in the BCS channel is
logarithmically enhanced, and the coupling itself runs double
logarithmically, in agreement with~\cite{Dzialosh,Gonzetal}.

\appendix*

\section{One-loop beta function calculations}

We consider only the spin-singlet interaction. The interaction part of
the Lagrangian is
\begin{equation} \label{interaction}
  \frac{g}{4} \epsilon^{ab} \epsilon^{cd}
  \psi_{a}^{\dagger} \psi_{b}^{\dagger} \psi_{c} \psi_{d}.
\end{equation}
The tree-level four-point amplitude is
\begin{equation} \label{tree}
  S_{abcd} g,
\end{equation}
where
\begin{equation} \label{sabcd}
  S_{abcd} = \epsilon_{ab} \epsilon_{cd}
\end{equation}
is the spin structure of the amplitude. Momentum conservation means
the coupling is a function of three momenta (six real variables)
$g(p_{1},p_{2},p_{3})$ which we take to be symmetric under
$p_{1}\leftrightarrow{}p_{2}$ and
$p_{3}\leftrightarrow{}p_{1}+p_{2}-p_{3}$.

The one-loop four-point amplitude has contributions from the $s$, $t$,
and $u$ channels. The $s$ channel has the same spin-singlet structure
as the tree-level amplitude, but the $t$ and $u$ channels generically
contain both spin-singlet and spin-triplet contributions. To simplify
this analysis, we ignore the spin-triplet contributions entirely. Then
the amplitude takes the form
\begin{equation} \label{one-loop}
  \amp_{abcd} = S_{abcd}(\amps + \ampt + \amp_{\rm{U}}).
\end{equation}
The $u$-channel amplitude follows from the $t$-channel amplitude via
exchange of the two outgoing momenta. Besides the momentum dependence
of the coupling, the $s$- and $t$-channel diagrams only depend on the
external momenta through $K=p_{1}+p_{2}$ and $Q=p_{1}-p_{3}$
respectively.

After performing the energy integrals via contours, changing
coordinates to $p_{\pm}=p_{x}\pm{}p_{y}$, and assuming time-reversal
invariance for the dispersion,
\begin{multline} \label{amps}
  \amps
  = -\frac{1}{8 \pi^{2}} \int d^{2}k
   \frac{\theta(\ve(k)) \theta(\ve(K-k))
     - \theta(-\ve(k)) \theta(-\ve(K-k))}
        {\ve(k) + \ve(K-k) - E - i \eps ~\sign{\ve(k)}}\\
        \times g(p_{1}, p_{2}, k) g(k, K-k, p_{3}) f(k) f(K - k).
\end{multline}
$E$ is the net energy of the external particles. $f$ contains all
information regarding the cutoffs:
\begin{equation} \label{cutoffs}
  f(k) =
  \theta(\la-|\ve(k)|) \theta(\V - |k_{+}|) \theta(\V - |k_{-}|).
\end{equation}
Similarly, the $t$-channel amplitude is
\begin{multline} \label{ampt}
  \ampt
  = \frac{1}{16 \pi^{2}} \int d^{2}k
   \frac{\theta(\ve(k)) \theta(-\ve(k+Q))
     - \theta(-\ve(k)) \theta(\ve(k+Q))}
        {\ve(k) - \ve(k+Q) + E_{\mathrm{T}} - i \eps ~\sign{\ve(k)}}\\
        \times g(p_{1}, k, p_{3}) g(k+Q, p_{2}, k) f(k) f(k + Q).
\end{multline}
$E_{\mathrm{T}}$ is the transfer energy of the external particles. The
extra factor of $1/2$ arises from isolating the spin-singlet
contribution.

We are interested in the beta function for $g$. Taking the logarithmic
derivative with respect to $\la$ yields
\begin{equation} \label{d-amps-simp}
  \logd{\amps}{\la}
  = -\frac{1}{4 \pi^{2}} (I_{S+} + I_{S-}),
\end{equation}
\begin{equation} \label{d-ampt-simpt}
  \logd{\ampt}{\la}
  = \frac{1}{8 \pi^{2}} (I_{T+} + I_{T-}),
\end{equation}
where
\begin{multline} \label{i-plus}
  I_{S\pm}
  \equiv \pm \la \int d^{2}k
  \frac{\delta(\la \mp \ve(k)) \theta(\pm\ve(K-k))
    \theta(\V - |k_{+}|) \theta(\V - |k_{-}|) f(k - K)}
       {\ve(k) + \ve(K-k)}\\
       \times g(p_{1}, p_{2}, k) g(k, K-k, p_{3}),
\end{multline}
\begin{multline} \label{i1}
  I_{T\pm} \equiv \pm\frac{1}{2} \la \int d^{2}k
  \frac{\delta(\la \mp \ve(k)) \theta(\mp\ve(k+Q))
    \theta(\V - |k_{+}|) \theta(\V - |k_{-}|) f(k+Q)}
       {\ve(k) - \ve(k+Q)}\\
       \times [g(p_{1}, k, p_{3}) g(k+Q, p_{2}, k)
       + g(p_{1}, -k-Q, p_{3}) g(-k, p_{2}, -k-Q)].
\end{multline}
We have dropped $E$ and $E_{\mathrm{T}}$ because they lead to
power-suppressed terms in the beta function. The remaining integrals
are similar to each other. They involve integrating over the
one-dimensional space where one of the particles in the loop has
$\ve=\pm\la$ and the other has either the same sign for $\ve$ (for the
$s$ channel) or the opposite sign (for the $t$ channel).

Define $P$ to be equal to $K$ for the $s$-channel diagram and $-Q$ for
the $t$-channel diagram. We exploit the $O(1,1)$ invariance of the
dispersion to replace $P$ in by
$\tilde{P}=\sqrt{|\ve(P)|}(\sign{P_{+}},~\sign{P_{-}})$ in each of the
integrals by changing variables:
\begin{equation} \label{so11+}
  k_{+} = k_{+}'/\eta,
\end{equation}
\begin{equation} \label{so11-}
  k_{-} = \eta k_{-}',
\end{equation}
with
\begin{equation} \label{eta-def}
  \eta \equiv \frac{\sqrt{|\ve(P)|}}{|P_{+}|}
  = \sqrt{\left| \frac{P_{-}}{P_{+}} \right|}.
\end{equation}
We may take $\eta$ to be less than one by exchanging $k_{+}$ and
$k_{-}$ if necessary. The step functions involving the rapidity cutoff
$\V$ are not invariant under this change of variables. In particular,
\begin{equation} \label{rapidity-cv-p}
  \theta(\V - |k_{+}|) \to \theta (\eta \V - |k_{+}|),
\end{equation}
\begin{equation} \label{rapidity-cv-m}
  \theta(\V - |k_{-}|) \to \theta (\V/\eta - |k_{-}|),
\end{equation}
\begin{equation} \label{rapidity-cv-kp}
  \theta(\V - |k_{+} - P_{+}|)
  \to \theta (\eta \V - |k_{+} - \sqrt{|\ve(P)|}~\sign{P_{+}}|),
\end{equation}
\begin{equation} \label{rapidity-cv-km}
  \theta(\V - |k_{-} - P_{-}|)
  \to \theta (\V/\eta - |k_{-} - \sqrt{|\ve(P)|}~\sign{P_{-}}|).
\end{equation}
These set the limits of integration on the remaining $k_{+}$ integrals
(once we have performed the $k_{-}$ integrals with the delta function)
if the energy constraints do not set stricter limits. With that in
mind, let us first analyze the limits in the absence of a rapidity
cutoff.

\subsection{Integration limits}\label{sec:integration-limits}

We can write generic expressions for the various possible integration
limits in each of the four remaining integrals. As before, take $P$ to
be either $K$ or $-Q$. Define
\begin{equation} \label{s-pm}
  s_{\pm} = \sign{P_{\pm}},
\end{equation}
\begin{equation} \label{s-lambda}
  s_{k} =
  \begin{cases}
    1 &\mathrm{~for~} I_{S+}, I_{T+},\\
    -1 &\mathrm{~for~} I_{S-}, I_{T-},
  \end{cases}
\end{equation}
\begin{equation} \label{s-ve}
  s_{p} =
  \begin{cases}
    1 &\mathrm{~for~} I_{S+}, I_{T-},\\
    -1 &\mathrm{~for~} I_{S-}, I_{T+}.
  \end{cases}
\end{equation}

The remaining $k_{+}$ integrals have limits at
\begin{equation} \label{C-intersect}
  \lambda_{A} \equiv s_{+} \sqrt{|\ve(P)|},
\end{equation}
\begin{equation} \label{D-intesect}
  \lambda_{B} \equiv s_{-} s_{k} \frac{\la}{\sqrt{|\ve(P)|}}.
\end{equation}
Whenever the quantities
\begin{multline} \label{hyperbola-intersect}
  \lambda_{\pm}
  \equiv \frac{1}{2} \sqrt{|\ve(P)|}
  \left(s_{+} + s_{-} (s_{k} - s_{p}) \frac{\la}{|\ve(P)|} \right. \\
  \left. \pm \sqrt{1
    - 2 s_{+} s_{-} (s_{k} + s_{p}) \frac{\la}{|\ve(P)|}
    + {(s_{k} - s_{p})}^{2} \frac{\la^{2}}{{\ve(P)}^{2}}} \right)
\end{multline}
are purely real, the remaining integrals also have limits at
$\lambda_{\pm}$. In that case the integration region splits into two
disjoint pieces.

For $I_{S+}$ and $I_{S-}$, $s_{k}=s_{p}$ and
Eq.~\eqref{hyperbola-intersect} simplifies to
\begin{equation} \label{hyperbola-intersect-pm}
  \lambda_{\pm} = \frac{1}{2} \sqrt{|\ve(P)|}
  \left(s_{+}
  \pm \sqrt{1 - 4 s_{+} s_{-} s_{k} \frac{\la}{|\ve(P)|}} \right).
\end{equation}
For $I_{T+}$ and $I_{T-}$, $s_{p}=-s_{k}$ and this simplifies to
\begin{equation} \label{hyperbola-intersect-12}
  \lambda_{\pm} = \frac{1}{2} \sqrt{|\ve(P)|}
  \left(s_{+} + 2 s_{-} s_{k} \frac{\la}{|\ve(P)|}
  \pm \sqrt{1 + \frac{4 \la^{2}}{{\ve(P)}^{2}}} \right).
\end{equation}
Therefore the $t$-channel integrals always split into two pieces. The
$s$-channel integrals split unless
\begin{equation} \label{s-channel-non-intersect}
  |\ve(K)| \le 4 \la,
\end{equation}
in which case $I_{S\sign{\ve(K)}}$ is over a single contiguous
region bounded by $\lambda_{A}$ and $\lambda_{B}$. In that case,
\begin{equation} \label{i-sign-ek}
  I_{S\sign{\ve(K)}}
  = \int_{\min{(\lambda_{A},\lambda_{B})}}^{\max{(\lambda_{A}, \lambda_{B})}} \dots
\end{equation}
We will see that this integral (and only this one) is generally
divergent as $\ve(K)\to0$, and that this divergence is cured by the
rapidity cutoff.

For the remaining three integrals (and for $I_{S\sign{\ve(K)}}$
if~\eqref{s-channel-non-intersect} is not satisfied), the integration
regions are bounded on one side by either $\lambda_{A}$ or
$\lambda_{B}$ and on the other by either $\lambda_{+}$ or
$\lambda_{-}$. The remaining integrals take the form
\begin{equation} \label{i-form}
  I = \left(
  \int_{\lambda_{1}}^{\lambda_{2}}
  + \int_{\lambda_{3}}^{\lambda_{4}} \right) dk_{+} \dots
\end{equation}
where $\lambda_{1}$ through $\lambda_{4}$ are the limits sorted in
ascending order.

\subsection{Rapidity limits}

At this point, let us simplify the discussion by taking $g$ to be a
momentum-independent constant. We will find that this assumption is
not consistent, because the beta function depends on the momentum.
Suppressing the integration limits, the remaining integrals take the
following form:
\begin{equation} \label{generic-i}
  I(P_{+}, P_{-}, s_{k}, s_{p})
  = g^{2} \int d^{2}k \frac{\delta(\ve(k) - s_{k}\la)
  \theta(s_{p} \ve(k-P)) F(k, P)}{1 + s_{p} \frac{\ve(k-P)}{\la}},
\end{equation}
where
\begin{equation} \label{cap-f}
  F(k, P)= \theta(\V - |k_{+}|) \theta(\V - |k_{-}|) f(k - P).
\end{equation}
Note that the step function constrains the value of the denominator to
be between 1 and 2 throughout the integration region, so all of the
integrals are nonnegative.

Under $P_{+}\to-P_{+}$ or $P_{-}\to-P_{-}$,
$I_{S+}\leftrightarrow~I_{S-}$ and $I_{T+}\leftrightarrow~I_{T-}$.
Thus we can take both components of $K$ and $Q$ to be positive without
loss of generality. This simplification would have held earlier if we
had assumed the coupling function obeys particle-hole symmetry, but
this symmetry is generically broken by the NVH region.

We can now find the effect of the rapidity cutoff on the integration
limits for the various integrals. The lower limit on $k_{+}$ imposed
by the rapidity cutoff for $I_{S+}$ and $I_{S-}$ is
\begin{equation} \label{lower-rapid-+}
  \lambda_{R1} = \eta \max{\left( \frac{\la}{\V}, P_{+}-\V\right)},
\end{equation}
where $\eta=\sqrt{|P_{-}/P_{+}|}$. The upper limit is
\begin{equation} \label{upper-rapid-+}
  \lambda_{R2} =
  \begin{cases}
    \eta \V, &P_{-}\le\V \\
    \eta \min{\left( \V, \frac{\la}{P_{-}-\V}, \right)}, &P_{-}>\V.
  \end{cases}
\end{equation}
$\lambda_{R1}$/$\lambda_{R2}$ replaces the lower/upper limits
in~\eqref{i-sign-ek} or~\eqref{i-form} when it is within either
integration region. Alternatively, if it is less than/greater than
both limits in one of the integrals, the integral is set to zero.

For $I_{S-}$ and $I_{T-}$, one of the integration regions has negative
$k_{+}$ and the other has positive $k_{+}$. There are four possible
rapidity limits:
\begin{align}
  \lambda_{R3} &= \eta (P_{+} - \V), \label{lr3} \\
  \lambda_{R4} &= -\frac{\eta \la}{\V}, \label{lr4} \\
  \lambda_{R5} &= \frac{\eta \la}{\V - P_{-}}, \label{lr5} \\
  \lambda_{R6} &= \eta \V. \label{lr6}
\end{align}
$\lambda_{R3}$/$\lambda_{R4}$ replace lower/upper limit for the
negative integration region and $\lambda_{R5}$/$\lambda_{R6}$ replace
the lower/upper limit for the positive region, or they set the
appropriate integrals to zero, acting in a manner analogous to that
described above for $\lambda_{R1}$ and $\lambda_{R2}$.

\subsection{Indefinite integrals}

Evaluating the
delta function in~\eqref{generic-i} and changing variables to
$x=\frac{k_{+}}{\sqrt{\la}}$ yields
\begin{equation} \label{generic-i-x}
  I = g^{2} \int \frac{dx}{|x|}
  \frac{1}{1 + s_{k} - s_{p} \alpha (x + s_{k}/x - \alpha)},
\end{equation}
where
\begin{equation} \label{alpha-beta}
  \alpha \equiv \sqrt{\frac{|\ve(P)|}{\la}}.
\end{equation}
We can directly compute the indefinite integrals for $I_{S\pm}$ and
$I_{T\pm}$ as long as we make use of the restrictions on the
integration limits implied by Section~\ref{sec:integration-limits}. We
find
\begin{equation} \label{i-p-anti}
  I_{S+}(x) = \frac{g^{2}}{\sqrt{\alpha^{4} + 4}}
  \log{\left|
    \frac{-2-\alpha^2+\sqrt{\alpha^4 + 4}+2 \alpha x}{
      2+\alpha^2+\sqrt{\alpha^4 + 4}-2 \alpha x}
    \right|},
\end{equation}
\begin{equation} \label{i-m-anti}
  I_{S-}(x) = \frac{g^{2} \sign{x}}{\sqrt{\alpha^{4} + 4}}
  \log{\left|
    \frac{-2+\alpha^2
      +\sqrt{\alpha^4 + 4}-2 \alpha x}{2
      -\alpha^2+\sqrt{\alpha^4 + 4}+2 \alpha x}
    \right|}.
\end{equation}
For $I_{T+}$, the appropriate indefinite integral depends on the
magnitude of $\alpha$, or in other words on the relative size of
$|\ve(Q)|$ and $\la$:
\begin{equation} \label{i-1-anti}
  I_{T+}(x) =
  \begin{cases}
    -\frac{2 g^{2}}{\alpha \sqrt{4 - \alpha^{2}}}
    \arctan{\left(
      \frac{\alpha - 2 x}{\sqrt{4 - \alpha^{2}}}\right)},
    &|\ve(Q)|<4\la,\\
    \frac{g^{2}}{\alpha \sqrt{\alpha^{2} - 4}}
    \log{\left|
      \frac{\alpha+\sqrt{\alpha^2-4}-2 x}{
        -\alpha+\sqrt{\alpha^2-4}+2 x} \right|},  &|\ve(Q)|>4\la.
  \end{cases}
\end{equation}
Finally,
\begin{equation} \label{i-2-anti}
  I_{T-}(x) = \frac{g^{2} \sign{x}}{\alpha \sqrt{\alpha^{2} + 4}}
  \log{\left| \frac{-\alpha+\sqrt{\alpha^2-4}+2 x}{
      \alpha+\sqrt{\alpha^2-4}-2 x} \right|}.
\end{equation}

\subsection{Collinear-anticollinear limit}

Consider scattering between generic collinear and anticollinear
particles. In this case, $|\ve(P)|\gg4\la$, so $\alpha\gg2$. In the
$\alpha\to\infty$ limit,
\begin{equation} \label{root-large-a}
  \frac{1}{\alpha \sqrt{\alpha^{2} + 4}}
  \approx \frac{1}{\alpha \sqrt{\alpha^{2} - 4}}
  \approx \frac{1}{\sqrt{\alpha^{4} + 4}}
  \to \frac{1}{\alpha^{2}} = \frac{\la}{|\ve(P)|}.
\end{equation}
This suggests the beta function is suppressed by $\la/|\ve(P)|$ for
collinear-anticollinear scattering, although we must also check the
behavior of the log functions.

When $a>2$, there are always two disjoint integration regions.
Exchanging $k_{+}$ and $k_{-}$ exchanges the two regions, so when we
ignore the rapidity cutoff they must have the same value. Evaluating
the integral between
\begin{equation} \label{lambda+}
  \frac{\lambda_{+}}{\sqrt{\la}} =
  \frac{1}{2} \alpha
  \left(1 + (s_{k} - s_{p}) \frac{1}{\alpha^{2}} + \sqrt{1
    - 2 (s_{k} + s_{p}) \frac{1}{\alpha^{2}}
    + {(s_{k} - s_{p})}^{2} \frac{1}{\alpha^{4}}} \right)
\end{equation}
and
\begin{equation} \label{lambda-b}
  \frac{\lambda_{A}}{\sqrt{\la}} = \alpha,
\end{equation}
reversing the order if $\lambda_{+}>\lambda_{A}$, and taking the
$\alpha\to\infty$ limit yields the same result for each integral:
\begin{equation} \label{collinear-result}
  I = (2 \log{2}) \frac{\la}{|\ve(P)|} g^{2}
  + \bo\left( \frac{\la^{2}}{{\ve(P)}^{2}} \right)
\end{equation}
for collinear-anticollinear scattering. All one-loop contributions to
the beta function are therefore power suppressed in this limit. This
remains true when we include the rapidity cutoff, since it can only
reduce the size of the integration region. Furthermore, such
interactions continue to be power suppressed after we drop the
assumption of a momentum-independent coupling, since the integration
region always shrinks to zero size as $\ve(P)$ becomes large.

\begin{figure}
  \epsfxsize=8cm\epsfbox{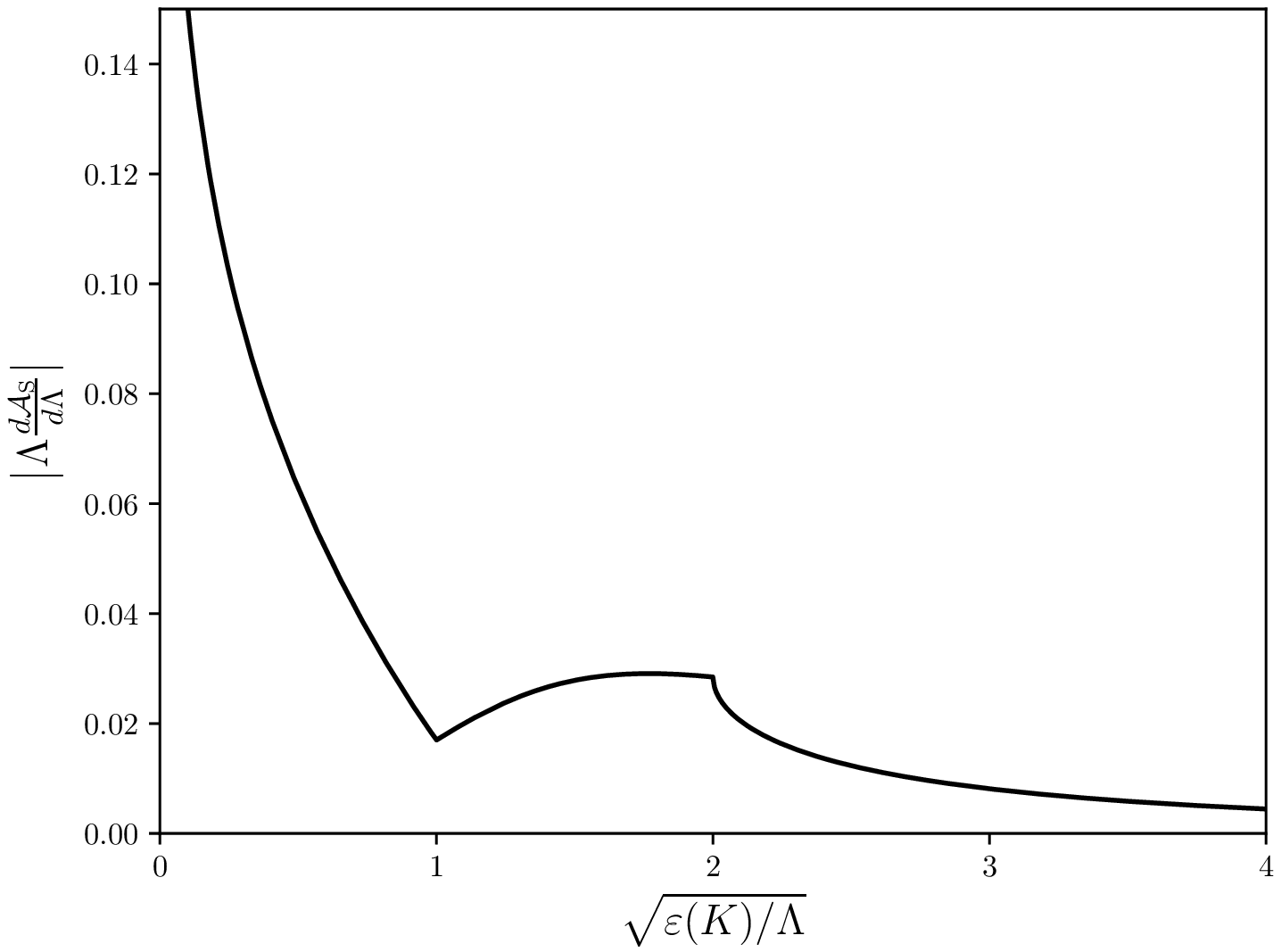}
  \epsfxsize=8cm\epsfbox{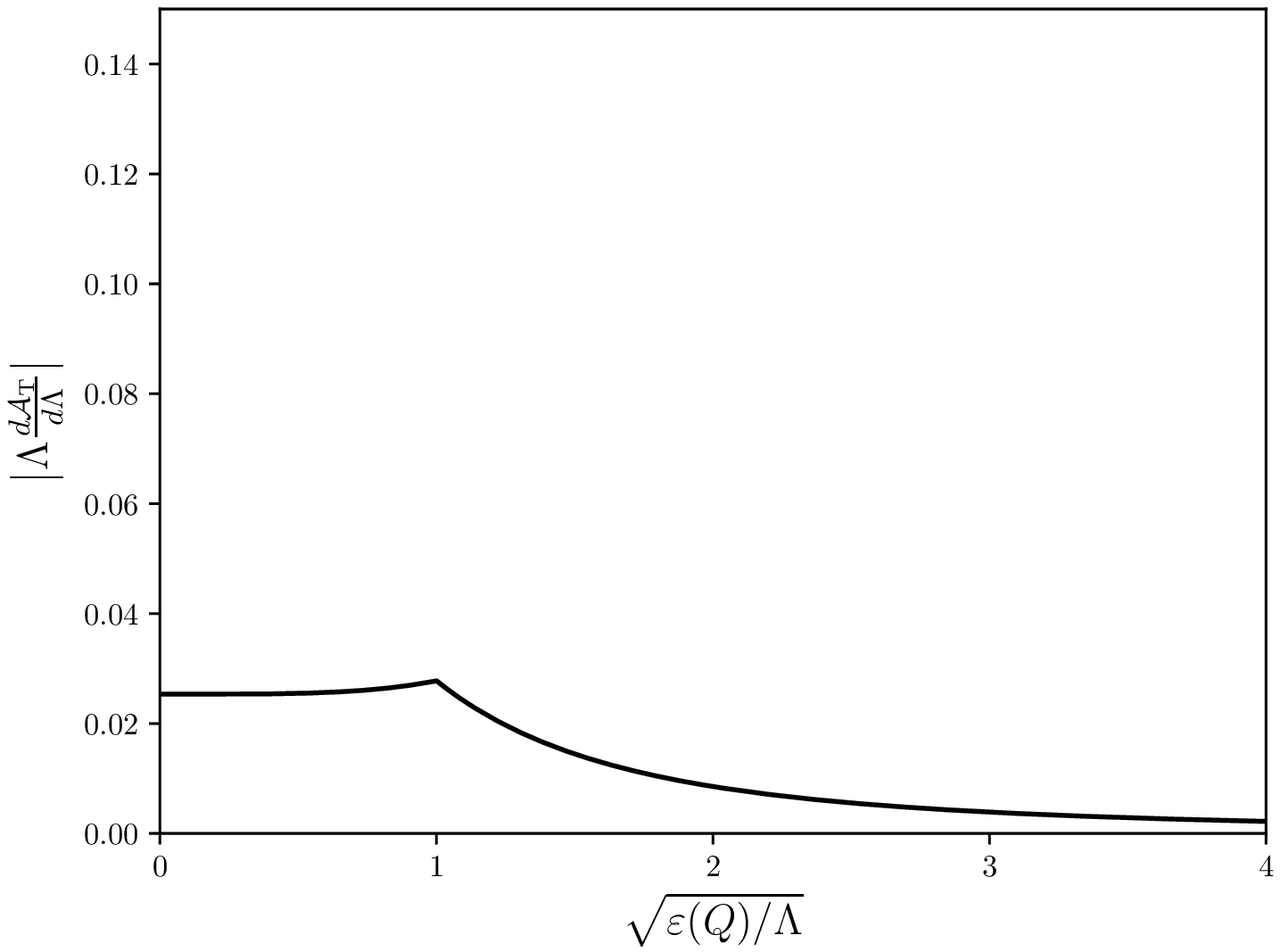}
  \caption{The log derivatives of the $s$- and $t$-channel diagrams in
    units of $g^{2}$ with no rapidity cutoff for a Fermi surface with
    a van Hove singularity. We assume a constant
    coupling.\label{fig:vh-betas}}
\end{figure}
\begin{figure}
  \epsfxsize=8cm\epsfbox{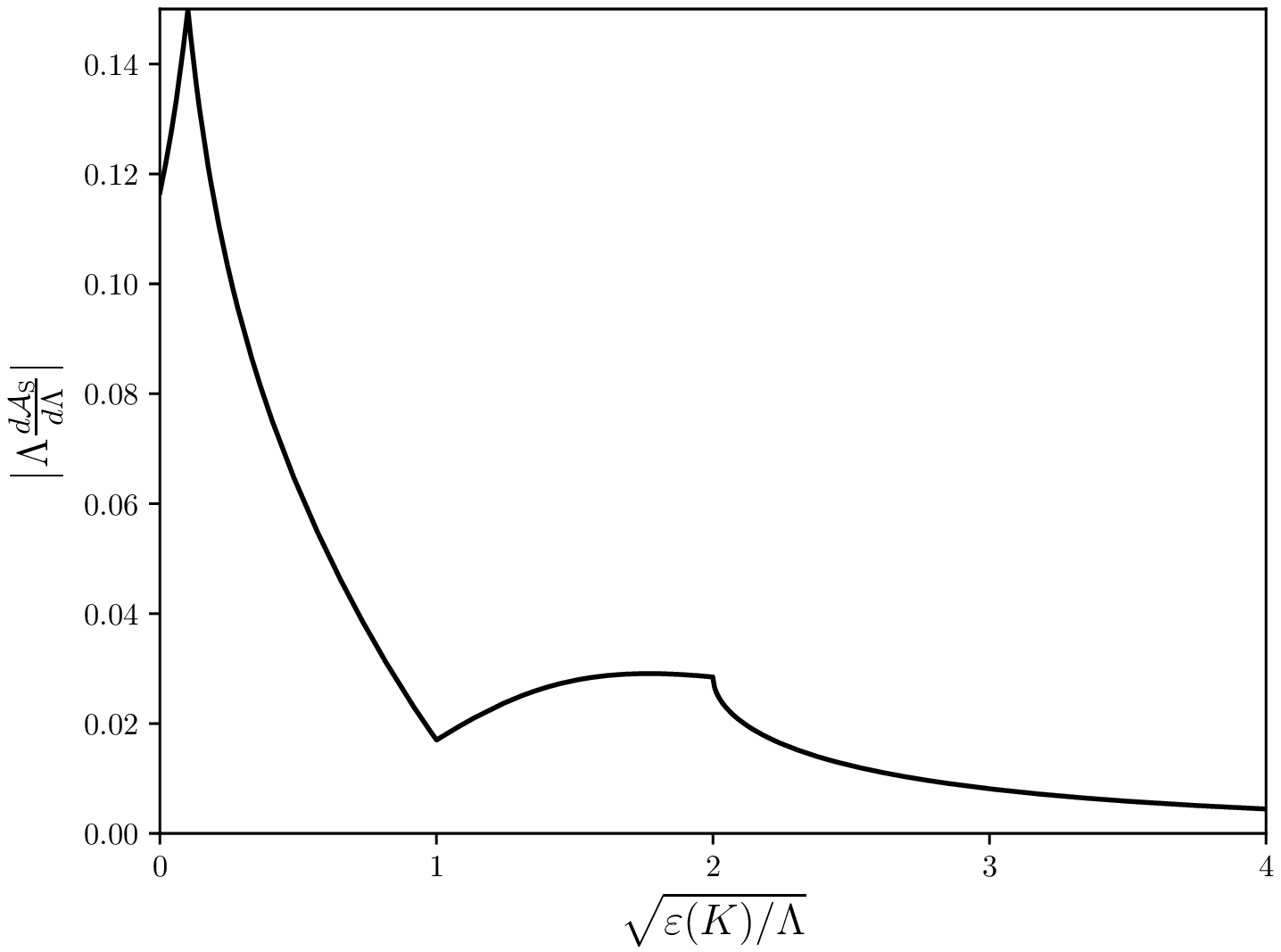}
  \epsfxsize=8cm\epsfbox{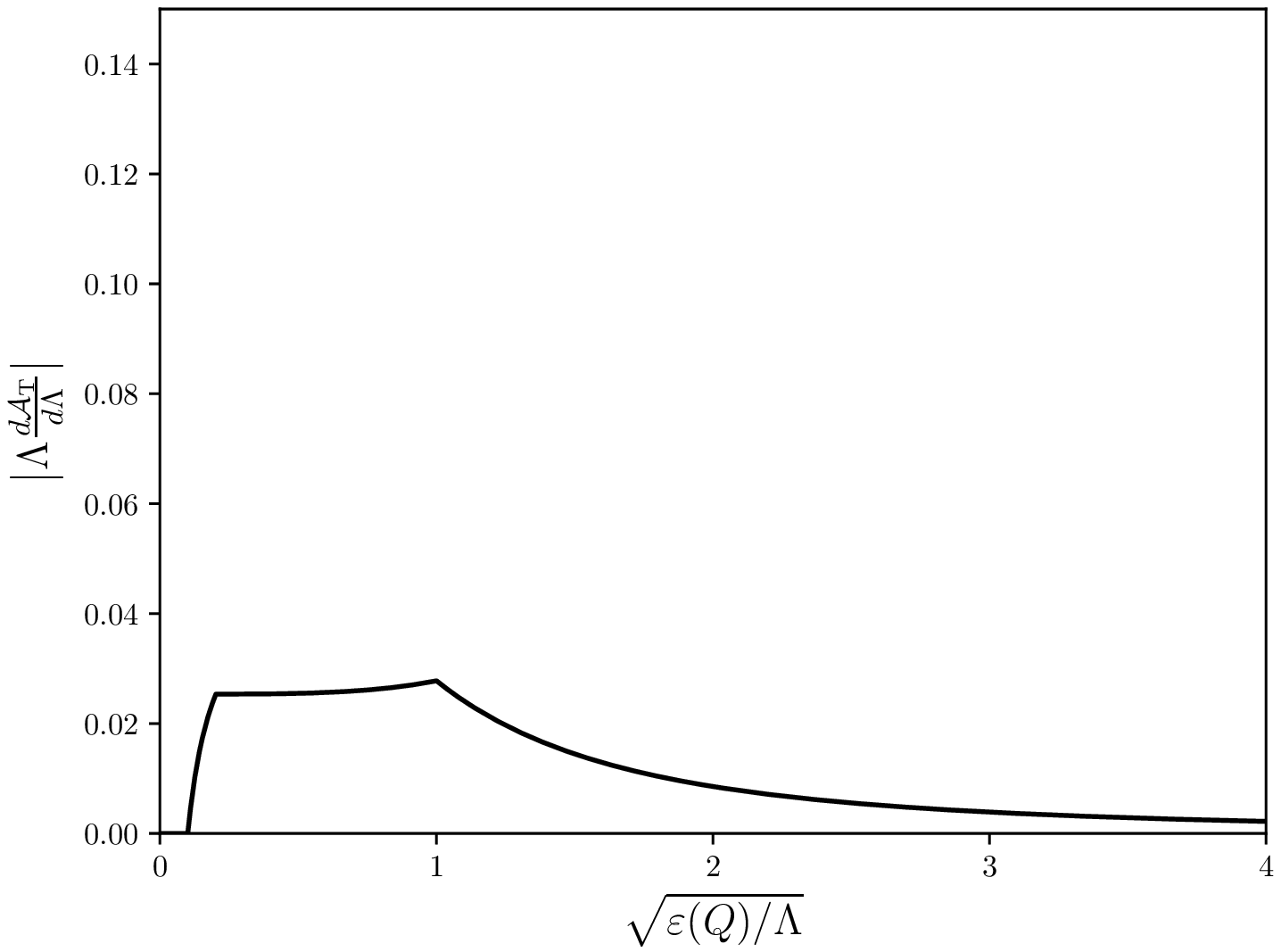}
  \caption{Plots demonstrating how the rapidity cutoff modifies
    Fig.~\eqref{fig:vh-betas}.\label{fig:vh-betas-rapidity}}
\end{figure}

\subsection{Collinear limit}

Consider the scenario where all scattered particles are restricted to
the collinear region and assume the external momenta $p_{i+}$ and
their sums/differences ($K_{+}$, $Q_{+}$, and $Q_{+}'$) are all order
$\V$ and much larger than $\sqrt{\la}$. Furthermore, assume the
scattered particles have energies well below the cutoff, so
$\ve(p)\ll\la$. Together, these imply that the perpendicular
components of momenta are small:
\begin{equation} \label{small-components}
  p_{i-} = \frac{\ve(p_{i})}{p_{i+}}
  = \bo\left( \frac{\ve(p_{i})}{\V} \right)
  \ll \frac{\la}{\V}.
\end{equation}
The following results also hold, with appropriate modifications, if
all momenta lie in the anticollinear region.

In this limit, only the rapidity cutoff on the collinear components of
momenta comes into play. Furthermore,
\begin{equation} \label{ve-limit-collinear}
  \ve(K) = \ve(p_{1})\left(1 + \frac{p_{2+}}{p_{1+}}\right)
  + \ve(p_{2})\left(1 + \frac{p_{1+}}{p_{2+}}\right)
  \ll \la
\end{equation}
since we have assumed that the collinear components of the incoming
particles are all of the same order. As a result,
\begin{equation} \label{alpha-small-collinear}
  \alpha = \sqrt{\frac{\ve(K)}{\la}} \ll 1.
\end{equation}
Similar statements hold for the $t$- and $u$-channel contributions.

After a change of variables, the integration limits for both the
$I_{S+}$ and $I_{S-}$ integrals are $\sqrt{|\ve(K)/\la|}=\alpha$ and
$\eta\V/\sqrt{\la}=\alpha\V/|K_{+}|$. Only one of the integration
regions for $I_{S-}$ remains after we impose the rapidity cutoff.
Substituting these limits into the indefinite integrals and taking the
small-$\alpha$ limit yields
\begin{equation} \label{i-p-m-collinear}
  I_{S+} \approx I_{S-}
  \to \frac{g^{2}}{2} \log{\left(\frac{2\V}{K_{+}} - 1\right)}.
\end{equation}

For the $I_{T+}$ and $I_{T-}$ integrals, only one of the two
integration regions remains. The limits are
\begin{equation} \label{lambda-plus-t}
  \frac{\lambda_{\mp}}{\sqrt{\la}}
  = \frac{1}{2} \alpha
  \left(1 \pm \frac{2}{\alpha^{2}}
  \mp \sqrt{1 + \frac{4}{\alpha^{4}}}\right),
\end{equation}
with $\lambda_{-}$ for $I_{T+}$ and $\lambda_{+}$ for $I_{T-}$,
and $\lambda_{A}/\sqrt{\la}=\alpha$. Substituting these into the
appropriate indefinite integrals gives
\begin{equation} \label{i-1-2-collinear}
  I_{T+} \approx I_{T-}
  \to \frac{g^{2}}{2}
\end{equation}
in the collinear limit.

There are two important features of~\eqref{i-p-m-collinear}
and~\eqref{i-1-2-collinear}. First, the contributions from the $s$,
$t$, and $u$ channels will all be order $g^{2}$. Second, the integrals
are independent of the small (anticollinear) components of the
external momenta. These conclusions do not depend of our assumption of
a momentum-independent coupling. Backtracking through our derivation
and restoring the momentum dependence
yields~\eqref{collinear-beta-final}.

\subsection{Forward scattering}

If both components of $Q$ are smaller than $\la/\V$, both integration
regions for $I_{T+}$ and $I_{T-}$ shrink to zero size. Thus, the
$t$-channel contribution to the beta function disappears in the
forward-scattering limit in the presence of a rapidity cutoff; see
Fig.~\ref{fig:vh-betas-rapidity}. This is analogous to the situation
discussed in~\cite{Shankar}, where the forward scattering function
makes no contribution to the beta functions for a round Fermi surface.
As in the case of a round Fermi surface, there is a sharp change in
the contribution to the beta function once $\ve(Q)$ exceeds a
threshold; compare Fig.~\ref{fig:round} and
Fig.~\ref{fig:vh-betas-rapidity}.

\subsection{BCS limit}

Consider $I_{S+}$ in the $K_{\pm}\to0$ limit. Since $\ve(K)<4\la$,
there is a single contiguous integration region, bounded by
$\lambda_{A}=\sqrt{|\ve(K)|}$ and $\lambda_{B}=\sqrt{\la/|\ve(K)|}$.
The extent of this region diverges as we lower $\ve(K)$. We find
\begin{equation} \label{bcs+}
  I_{S+} = \frac{g^{2}}{\sqrt{\alpha^4+4}}
  \log{\left( \frac{(-2+\alpha^2+\sqrt{\alpha^4+4})      
      (\alpha^2+\sqrt{\alpha^4+4})}
    {(-\alpha^2+\sqrt{\alpha^4+4})
      (2-\alpha^2+\sqrt{\alpha^4+4})} \right)}.
\end{equation}
Taking the small $\alpha$ limit yields
\begin{equation} \label{bcs+-diverge}
  I_{S+} \to \frac{1}{2} g^{2} \log{\frac{4 \la}{\ve(K)}},
\end{equation}
which diverges at $\ve(K)=0$. This is the divergence that forced us to
introduce the rapidity regulator. $I_{S-}$ has the same value as
$I_{S+}$ in the small $\alpha$ limit.

Introducing the rapidity cutoff regulates the divergence. The rapidity
cutoff restricts the $I_{S+}$ integral to run from $\eta\sqrt{\la}/\V$
to $\eta\V/\sqrt{\la}$ and the $I_{S-}$ integral to run from $\eta$ to
$\eta\V/\sqrt{\la}$. Plugging these into~\eqref{i-p-anti}
and~\eqref{i-m-anti} and taking the $\alpha\to0$ limit yields
\begin{equation} \label{i-rapidity-a-limit}
  I_{\pm} \to \frac{1}{2} g^{2} \log{\frac{\V^{2}}{\la}},
\end{equation}
so
\begin{equation} \label{final-bcs}
  \logd{\amps}{\la}
  = -\frac{g}{4 \pi^{2}} \log{\frac{\V^{2}}{\la}}
\end{equation}
for back-to-back interactions.

\subsection{Generic BCS beta function}

\begin{figure}
  \begin{centering}    
    \epsfxsize=16cm\epsfbox{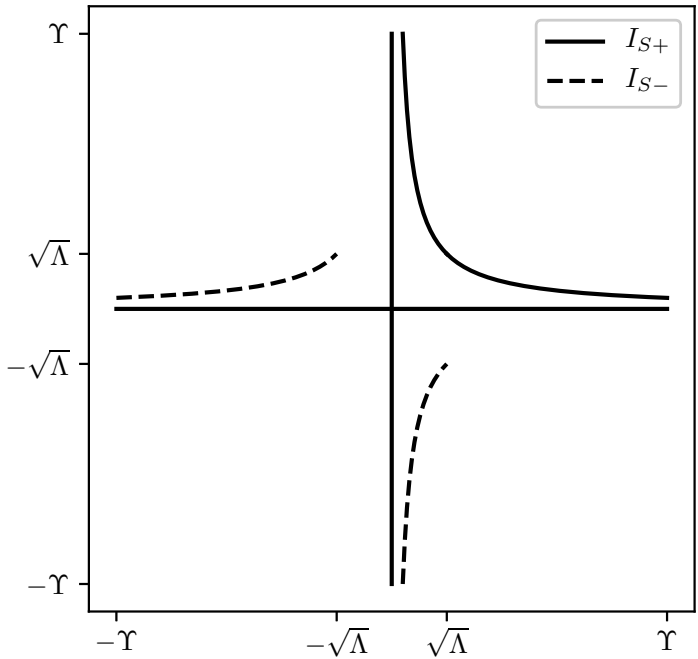}
    \caption{The integration regions for
      $I_{S\pm}$.\label{fig:k-infinitesimal}}
  \end{centering}
\end{figure}

The previous results indicate that we may take the coupling for fixed
ultrasoft net momentum (the BCS configuration) to be analytic in the
other momenta to leading-log order. Furthermore, we may drop all but
the $s$-channel diagram to this order. Parameterize the BCS coupling
$\gb(p_{1}, p_{3})$ in terms of one of the incoming momenta $p_{1}$
and one of the outgoing momenta $p_{3}$ at fixed ultrasoft $K$. The
log derivative of the amplitude is~\eqref{d-amps-simp}, with
\begin{equation} \label{i-plus-bcs}
  I_{S\pm}
  \equiv \pm \la \int d^{2}k
  \frac{\delta(\la \mp \ve(k)) \theta(\pm\ve(K-k))
     F(k, K)}{2 \ve(k) - E} \gb(p_{1}, k) \gb(k, p_{3})
\end{equation}
and $F(k, K)$ from~\eqref{cap-f}. Take the components of $K$ to be
positive but infinitesimal to avoid ambiguity from the definition of
the step functions. Eq.~\eqref{i-plus-bcs} receives contributions from
several one-dimensional regions of momentum space; see
Fig.~\ref{fig:k-infinitesimal}.

Call $I_{++}$ the contribution from the region with $k_{+}>k_{-}>0$.
Evaluating the $k_{-}$ integral with the delta function yields
\begin{equation} \label{i++}
  I_{++}
  = \frac{1}{2}
  \int_{\sqrt{\la}}^{\V} 
  \frac{d k}{k} \gb(p_{1}, k) \gb(k, p_{3})
\end{equation}
up to power-suppressed terms. Assume it is possible to expand the
coupling function in $k_{+}$ and $k_{-}$. The resulting expression for
$I_{++}$ will include terms of the form
\begin{equation} \label{expansion-term}
  \int_{\sqrt{\la}}^{\V} \frac{d k_{+}}{k_{+}}
  k_{+}^{m}{\left(\frac{\la}{k_{+}}\right)}^{n}
  \partial_{k_{+}}^{m} \partial_{k_{-}}^{n} [\gb(p_{1}, 0) \gb(0, p_{3})].
\end{equation}
The natural scale for the derivatives is ${(1/V_{F})}^{m+n}$. As a
result, terms with $m\ne{}n$ give at most order-one contributions to
the beta function. When $m=n$, terms in~\eqref{expansion-term} take
the form
\begin{equation} \label{k}
  {\left( \frac{\la}{V_{F}^{2}} \right)}^{n}
  \int_{\sqrt{\la}}^{\V} \frac{d k_{+}}{k_{+}} =
  \frac{1}{2} {\left( \frac{\la}{V_{F}^{2}} \right)}^{n}
  \log{\frac{\V^{2}}{\la}}.
\end{equation}
Since we assume $\la\ll{}V_{F}^{2}$, these are suppressed unless
$n=0$. The $n=m=0$ term is log enhanced, and the leading-log result is
therefore
\begin{equation} \label{leading}
  I_{++}
  = \frac{1}{4} \gb(p_{1}, 0) \gb(0, p_{3}) \log{\frac{\V^{2}}{\la}}.
\end{equation}

A similar analysis holds for each of the the terms in $I_{S+}+I_{S-}$.
Adding the NVH region cancels the $\V$ dependence. Finally, setting
the log derivative with respect to $\la$ of the sum of the tree-level
amplitude $\gb(p_{1}, p_{3})$ and the one-loop amplitude equal to zero
implies
\begin{equation} \label{leading-beta}
  \logd{\gb(p_{1}, p_{3})}{\la}
  = \frac{1}{4 \pi^{2}} \gb(p_{1}, 0) \gb(0, p_{3})
  \log{\frac{V_{F}^{2}}{\la}}.
\end{equation}

The beta function for the the coupling between modes in the vicinity
of the VH point, $\gb(0,0)$, is independent of the other couplings,
and the solution is
\begin{equation} \label{soft-solved}
  \gb(0, 0; \la)
    = \frac{\gb(0, 0, \la_0)}{1
      + \frac{\gb(0, 0, \la_0)}{8\pi^2}
      \left(\log^2 \frac{V_F^2}{\la}
      - \log^2\frac{V_F^2}{\la_0}\right)}.
\end{equation}
Using this, the beta function for $\gb(p_{1}, 0)$ becomes
\begin{equation} \label{collinear-soft-sub-soft}
  \la \frac{d \gb(p_{1}, 0)}{d \la}
  = \frac{1}{4 \pi^2} \frac{\gb(p_{1}, 0) \gb(0, 0, \la_0)}{1
    + \frac{\gb(0, 0, \la_0)}{8\pi^2}
    \left(\log^2 \frac{V_F^2}{\la}
    - \log^2\frac{V_F^2}{\la_0}\right)}
  \log{\frac{V_{F}^{2}}{\la}},
\end{equation}
with solution
\begin{equation} \label{collinear-soft-solved}
  \gb(p_{1}, 0; \la)
    = \frac{\gb(p_{1}, 0; \la_{0})}{1
      + \frac{\gb(0, 0, \la_0)}{8\pi^2}
      \left(\log^2 \frac{V_F^2}{\la}
      - \log^2\frac{V_F^2}{\la_0}\right)}.
\end{equation}
An analogous result holds for $\gb(0,p_{3};\la)$. Substituting these
into the beta function for $\gb(p_{1}, p_{3})$ and solving yields
\begin{equation} \label{collinear-solved}
  \gb(p_{1}, p_{3}; \la)
    = \gb(p_{1}, p_{3}; \la_{0}) - 
    \left( \frac{1}{8 \pi^2} \right)
    \frac{\gb(p_{1}, 0; \la_{0}) \gb(0, p_{3}; \la_{0})
      \left(\log^2 \frac{V_F^2}{\la}
      - \log^2\frac{V_F^2}{\la_0}\right)}{1 +
      \frac{\gb(0, 0; \la_{0})}{8\pi^2}
      \left(\log^2 \frac{V_F^2}{\la}
      - \log^2\frac{V_F^2}{\la_0}\right)}.
\end{equation}
We see that the expressions for $\gb(0,0;\la)$, $\gb(p_{1},0;\la)$,
and $\gb(0,p_{3};\la)$ are in fact special cases of this general
result.

\begin{acknowledgments}
This work was supported by the DOE contracts \uppercase{DE-SC0011632},
DE-FG02-92ER40701, \uppercase{DOE-ER}-40682-143, and
\uppercase{DE-AC02-6CH03000}. The authors gratefully acknowledge
helpful conversations with Joe Polchinski, Max Metlitskii, Lesik
Motrunich, and Ingmar Saberi. AK and TM are grateful to the Simons
Center for Geometry and Physics for hospitality during various stages
of this work. AK is also grateful to the Aspen Center for Physics, the
Kavli Institute for Physics and Mathematics of the Universe, and
Institut des Hautes Etudes Scientifiques for hospitality. IZR is
grateful to the Caltech theory group for hospitality and to the Moore
Foundation for support.
\end{acknowledgments}

\end{document}